\documentclass[lettersize,journal]{IEEEtran}
\usepackage{cite}
\usepackage{tikz}
\usepackage{tikzscale}
\usepackage{subcaption}
\usepackage{makecell}
\usepackage{xstring}
\usepackage{multirow}
\usepackage{amsmath,amssymb,amsfonts}
\usepackage[normalem]{ulem}

\usepackage{amsmath}
\usepackage{algorithm}
\usepackage{etoolbox}\AtBeginEnvironment{algorithmic}{\scriptsize}
\usepackage[noend]{algpseudocode}
\usepackage{hyperref}

\usetikzlibrary{
    calc,
	decorations.pathreplacing,
	positioning,
	shapes.misc,
	tikzmark,
}

\usepackage{enumitem}

\def\eg{e.g.,~}
\def\ie{i.e.,~}

\def\cf{cf.~}

\newcommand{\redfoot}[1]{{\color{red}#1}}

\newcommand{\blue}[1]{{\color{black}#1}}
\newcommand{\teal}[1]{{\color{black}#1}}

\begin{document}

\title{AutoML4ETC: Automated Neural Architecture Search for Real-World Encrypted Traffic Classification}

\author{
	\IEEEauthorblockN{
 	Navid Malekghaini\IEEEauthorrefmark{1}, Elham Akbari\IEEEauthorrefmark{1}, Mohammad A. Salahuddin\IEEEauthorrefmark{1},
 	 Noura Limam\IEEEauthorrefmark{1}, Raouf Boutaba\IEEEauthorrefmark{1},\\Bertrand Mathieu\IEEEauthorrefmark{2}, Stephanie Moteau\IEEEauthorrefmark{2}, and Stephane Tuffin\IEEEauthorrefmark{2}\\
 	}
 	\IEEEauthorblockA{
 	\IEEEauthorrefmark{1}David R. Cheriton School of Computer Science, University of Waterloo, Ontario, Canada\\
 	\texttt{\{nmalekgh, eakbaria, mohammad.salahuddin, noura.limam, rboutaba\}@uwaterloo.ca}
 	}
 	
  	\IEEEauthorblockA{
 	\IEEEauthorrefmark{2}Orange Labs, Lannion, France\\
 	\texttt{\{bertrand2.mathieu, stephanie.moteau, stephane.tuffin\}@orange.com}
 	}}

\maketitle

\begin{abstract}

    Deep learning (DL) has been successfully applied to encrypted network traffic classification in experimental settings. However, in production use, it has been shown that a DL classifier's  performance  inevitably decays over time. Re-training the model on newer datasets has been shown to only partially improve its performance. Manually re-tuning the model architecture to meet the performance expectations on newer datasets is time-consuming and requires domain expertise. We propose AutoML4ETC, a novel tool to automatically design efficient and high-performing neural architectures for encrypted traffic classification. We define a novel, powerful search space tailored specifically for the early classification of encrypted traffic using packet header bytes. We show that with different search strategies over our search space, AutoML4ETC generates neural architectures that outperform the state-of-the-art encrypted traffic classifiers on several datasets, including public benchmark datasets and real-world TLS and QUIC traffic collected from the Orange mobile network. In addition to being more accurate, AutoML4ETC’s architectures are significantly more efficient and lighter in terms of the number of parameters. Finally, we make AutoML4ETC publicly available for future research.\footnote{\redfoot{Paper has been accepted for publication in IEEE TNSM journal.}}
\end{abstract}


\section{Introduction}
\blue{
Traffic classification (TC) is an important task for the operation and management of computer networks. It is essential for traffic analysis and can be used for effective network planning, resource provisioning and allocation, providing differentiated Quality-of-Service (QoS), improving customer Quality of Experience (QoE), monitoring network security, etc. For instance, the Internet service provider (ISP) can use TC to identify different types of traffic in the network and to differentiate the services provided to them, \eg prioritizing certain types of traffic over others. Moreover, it could be used to guarantee service and resources to critical business services that are associated with a specific type of traffic. Another example would be the use of TC in Intrusion Detection Systems (IDS); with TC, an IDS could distinguish unknown network traffic from other types of traffic and take appropriate actions, such as blocking them. This example is an application of TC for anomaly detection. Due to the pervasive use of encryption nowadays, the bulk of Web-based applications communicate using Hypertext Transfer Protocol Secure (HTTPS), which relies on Transport Layer Security (TLS) to encrypt the application message, making TC a difficult task.
}

Deep learning (DL) has been widely employed for encrypted traffic classification (ETC) with performance exceeding traditional machine learning (ML) methods \cite{aceto2019mobile, lotfollahi2020deep, liu2019fs, seq2img, CACM, deeptor, FSTC}. However, the classification performance is known to vary based on the model architecture and the target dataset. The effect of the dataset on  model performance, in particular the distribution of data across classes, is especially important, as some distributions are easier to learn than others. For real-world ETC datasets, the data collection process, e.g., location of monitoring sensors in the network, duration of traffic capture, and employed filters, can affect the data distribution. For example, a dataset collected from a local area network is expected to have a different distribution from an ISP-level network dataset, as the larger the number of users in a network, the more diverse the generated traffic.

In 2020, we designed UWOrange~\cite{Baseline}, a DL-based service- and application-level encrypted traffic classifier with a novel tripartite architecture that out-performed state-of-the-art classifiers on Orange mobile traffic data as well as existing, publicly available datasets. However, as time went by, we observed that the performance of UWOrange on our ISP partner's newer traffic data was decaying. We thoroughly investigated this issue and reported in~\cite{IFIPPaper, COMNETPaper} how in a production setting, even when the data collection process is the same, state-of-the-art DL encrypted traffic classifiers (including UWOrange) are prone to performance decay. We highlighted the stark differences in state-of-the-art DL model performance across datasets, which was partly attributed to a change in the properties and statistical distribution of data over time, also known in the ML community as \emph{data drift}. 

Our findings suggested that in production, periodic model re-training over newer datasets was inevitable to alleviate the effect of data drift. However, re-training UWOrange on newer datasets did not prevent performance loss, which was particularly significant on specific datasets~\cite{IFIPPaper, COMNETPaper}. The decreasing performance of the re-trained classifier on newer datasets led us to assert that the architecture of UWOrange, initially designed to perform particularly well on baseline datasets, was no longer suitable for newer datasets. It became clear that in production, the hyper-parameters of the DL classifier, \eg learning rate, had to be re-tuned and the model architecture adapted by a domain expert on a regular basis, both of which are time-consuming and primarily based on trial-and-error.

When experimenting with UWOrange on Orange network data, it also became apparent that the use cases of interest required a more efficient, light-weight model architecture, capable of classifying traffic with \blue{just a few initial packets per flow (\ie early classification)} with high accuracy. These requirements, in addition to the challenges posed by data drift, i.e., constant, manual re-tuning or re-tweaking of the model architecture, motivated the need for an automated tool to design neural network architectures that are: (i) capable of classifying an encrypted flow with high accuracy from the first few packets, and (ii) lighter and more resource-effective models with comparable or better performance than the state of the art. Such a tool can be used to automatically produce new architectures adapted to newer datasets instead of manually re-tuning older ones.

Automating the process of finding the right architecture with the right hyper-parameters is a problem known to DL. In meta-learning, it was previously suggested to use  a supervisory neural network to learn the hyper-parameters of a subordinate neural network~\cite{scholarpedia, metalearnclassic1}. Recently, Neural Architecture Search (NAS) \cite{NASsurvey,NAS}, a sub-field of Automated Machine Learning (AutoML) \cite{automl_classic}, was introduced to automatically learn the best neural network architecture given a particular dataset. NAS arose from the extensive architectural engineering effort needed every time a new image classification dataset emerges. This closely aligns with the practical challenges pertaining to ETC. However, NAS requires choosing the building blocks of the architecture search space, which is not trivial and requires domain expertise. 

In this regard, our main contributions are as follows: 
\begin{itemize}

\item We propose AutoML4ETC, a novel resource-effective tool to automatically design efficient and
high-performing neural network architectures for ETC, given a target dataset. We show that in addition to being more accurate, the AutoML4ETC’s architectures are significantly more efficient and lighter than state-of-the-art encrypted traffic classifiers in terms of the number of parameters.

\item We define a novel, powerful architecture search
space tailored specifically for producing efficient encrypted traffic classifiers that leverage packet raw bytes. The building blocks of the search space were carefully selected after we exhaustively studied state-of-the-art architectures for ETC and extensively experimented with different architectures on real-world encrypted network traffic collected on the Orange mobile network. We show that the produced models achieve high classification accuracies from the first three Transport Layer Security (TLS) handshake packets and, for the first time in ETC literature, from the singleton \emph{ClientHello} packet of the Quick UDP Internet Connection (QUIC) protocol. This makes these classifiers suitable for early traffic classification.

\item We validate AutoML4ETC by extensively experimenting with real-world encrypted network traffic collected on the Orange mobile network. AutoML4ETC's architectures show superior performance both on QUIC and TLS traffic, with service-level  and application-level classification alike. We further experiment with publicly available datasets for the reproducibility of results and make the tool publicly available\footnote{\label{{foot:code}}\href{https://github.com/OrangeUW/AutoML4ETC}{https://github.com/OrangeUW/AutoML4ETC}} to advance the state of the art.
\end{itemize}

The remainder of the paper is organized as follows. Section~\ref{sec:related} describes the background and discusses related literature works. Section~\ref{sec:method} provides an exposition of the core components of  \textit{AutoML4ETC}. 
Section~\ref{sec:eval} evaluates AutoML4ETC's search space, search algorithms, and training strategies. It showcases the efficacy of the AutoML4ETC-generated architectures over public  and real-world datasets, and demonstrates their superior performance in comparison to state-of-the-art encrypted traffic classifiers. Finally, Section~\ref{sec:conclude} concludes the paper and instigates future directions. 
\section{Background and Related Works} 
\label{sec:related}

\subsection{Neural Architecture Search}

\begin{figure*}[tb]
\centerline{\includegraphics[width=\textwidth]{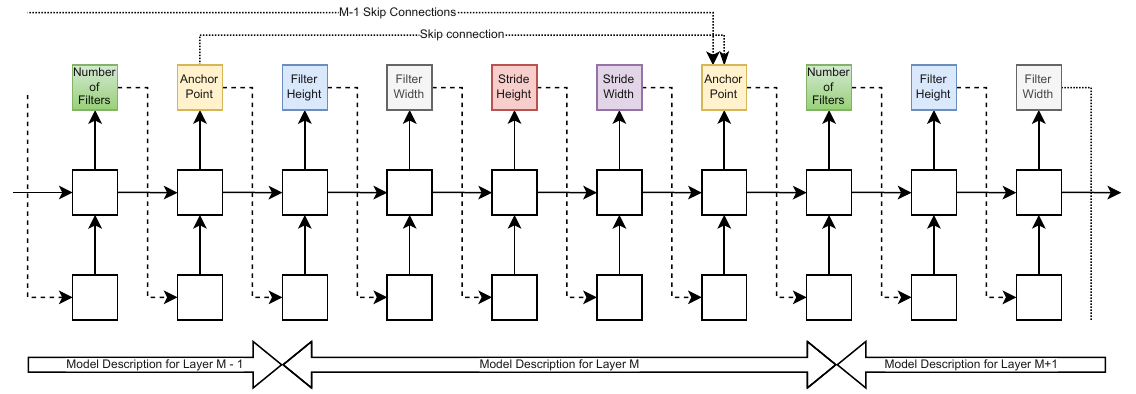}}
\caption{
Example CNN model descriptions and skip connections for each layer \cite{NAS}}
\label{fig:NAS-SEARCH}

\end{figure*}

NAS \cite{NAS} leverages a controller Recurrent Neural Network (RNN) to generate 
neural network architectures. 
Figure~\ref{fig:NAS-SEARCH} shows an architecture generated by NAS that consists of convolutional layers only, where a layer in the Convolutional Neural Network (CNN) is described by a sequence of tokens. Each token determines a separate characteristic of the convolutional layer, such as filter size and stride. The different compositions and permutations of the sequences determine the model architectures that can be generated, which represents the search space. 

\begin{figure}[tb]
\centerline{\includegraphics[width=\columnwidth]{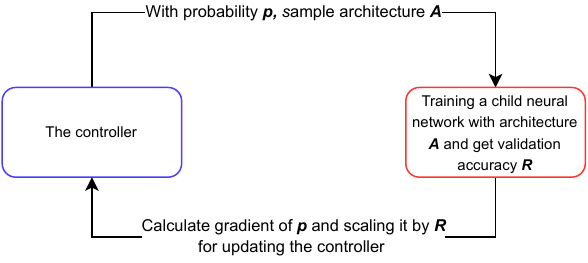}}
\caption{Overview of NAS with RL \cite{NAS}}
\label{fig:rl_rnn}

\end{figure}

In NAS, a controller RNN is trained using Reinforcement Learning (RL),
where the actions are the choice of tokens while the reward signal is the validation accuracy of the model, i.e., the sequence of tokens. RL consists of a series of trials, where a \textit{child model} is created by sampling the parameter values generated by the RNN at the end of each trial. As shown in Figure~\ref{fig:rl_rnn}, a sampled model is trained and evaluated on a dataset per trial to compute the reward, which makes NAS computationally expensive.

The RL algorithm described above operates within a space of possible sequences. This space is decided by the set of possible tokens that the controller RNN can generate at each time step. It is up to the domain expert to determine the set of possible tokens for RNN, which is similar to the set of words in the dictionary of a language generator. Researchers in \cite{NAS} proposed two different search spaces for creating both CNNs and RNNs. The authors increased the complexity of the convolutional models by introducing anchor points (\cf Figure~\ref{fig:NAS-SEARCH}) into the search space, which determines the probability of having skip connections between a layer and its previous layers. This allowed the architecture to contain branching or skip connections similar to the ones in ResNet \cite{Resnet}. Their results showed that the generated CNN models perform within an error rate of 1\% from the state-of-the-art image classifiers on the CIFAR-10 dataset \cite{cifar10}. The high performance is attributed to training 12,800 architectures using 800 GPUs for concurrent training, which makes their approach very resource-intensive. 

Authors in \cite{NAStoENAS} enhanced the set of tokens, \ie architecture building blocks that the RNN generates in \cite{NAS}. Their enhancement is based on the observation that state-of-the-art image classifier architectures have repeated network motifs, \ie small building blocks in the architecture's graph that are replicated. Their proposed search space consists of a sequence of \textit{Normal}  and \textit{Reduction} cells, in which only the Reduction cells reduce the size of the feature map. Between different architectures in the search space there is variation in the interior structure of the Normal and Reduction cells. Each cell is made up of a constant number of network motifs, where each motif consists of two inputs fed into two blocks that are aggregated. The types of blocks (\eg separable convolution, identity, 1x1 convolution) and the aggregation function (\eg add, concatenate) are determined by the controller RNN along with the connections between the motifs. Their method performed slightly better than the best recorded performance on CIFAR-10, with the added benefit of being transferable to the larger ImageNet dataset \cite{deng2009imagenet} despite the computational complexity of NAS. The authors leveraged a transfer learning approach to speed up child model training and promote transferability.

Both previous approaches suffer from high computational complexity. To address this problem, \cite{ENAS} improved NAS's time complexity by a factor of 1000 and achieved an error rate of within 0.3\% of NAS. \blue{To make this happen}, the authors employed parameter sharing among all child models, which is inspired by  transfer learning \cite{NAStoENAS} and multi-task learning \cite{multitask}. The authors named their approach Efficient Neural Architecture Search (ENAS). ENAS uses a more restrictive search space, where only the child models that can be represented by a directed acyclic graph are considered. Moreover, their micro search space is without non-separable convolutional blocks. The results of the micro search space are then compared to those of the search space in NAS
. Our proposed novel search space is inspired by the search space in ENAS.

The choice of the search algorithm has also been explored in NAS, where some works leveraged RL while others resorted to Evolutionary Algorithms (EAs). Outside the realm of NAS, \cite{MCTSTheory} proposed Monte Carlo Tree Search (MCTS) that extended the well-known Multi-armed Bandit technique in RL to tree-structured search spaces. This inspired an interesting approach in \cite{MCTS} that improved the controller by using MCTS to find the best architecture hyper-parameters. Using MCTS with Upper Confidence bound applied to Trees (UCT)
is known to balance exploitation and exploration in the searching process and overcome possible sub-optimal solutions. 
\teal{The main idea is to use MCTS to find the model's hyper-parameters in a layer-by-layer fashion in the child model descriptions. Selection, Expansion, Playout with simulation, and Backpropagation are the main steps of MCTS. To reliably estimate the search directions in MCTS, the child models are trained multiple times. The authors suggested using simulation to estimate the child model's accuracy, such that the child model is trained only once on the dataset, as opposed to multiple times, which saves on training time. The model's accuracy is then estimated by aggregating the training and simulation results.}

EAs are an alternate to RL for searching the neural architecture search space \cite{EAlarge, EA}. Authors in \cite{EAlarge} evolved an initial population of strings representing neural architectures by using a tournament selection algorithm, where after each pairwise comparison the worse individual dies and the better one mutates. The fitness of each string is determined by the respective architecture's validation accuracy after being trained on a dataset. Similar to \cite{ENAS}, researchers in \cite{EA} used an EA to search the NASNet search space. The authors used a tournament selection algorithm similar to \cite{EAlarge} and introduced the concept of aging to individuals. Comparing their algorithm to the RL baseline, they showed that their EA achieves a higher accuracy faster than the RL-based method, however, both methods converged to the same accuracy asymptote. The authors argued that the EA-based method is more relevant in larger search spaces where reaching the optimal solution may be resource intensive. We compare the EA algorithm in \cite{EA} to other search algorithms over our search space in Section \ref{sec:eval}.

\subsection{AutoML for Encrypted Traffic Classification}

\begin{figure*}[th]
\centerline{
\includegraphics[width=0.9\textwidth]{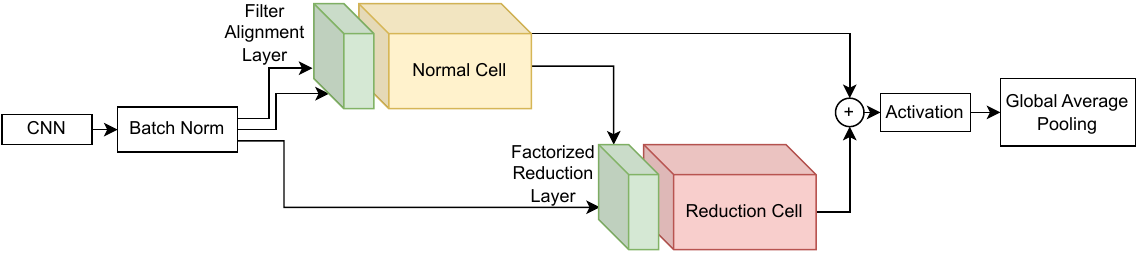}}
\caption{AutoML4ETC search space}
\label{fig:automl_packetraw_design}
\end{figure*}

\blue{

\blue{
A plethora of machine learning models have been suggested for traffic classification since the early 2000s. Numerous classical machine learning models including Naive Bayes, Bayesian Networks, k-Nearest Neighbor, and Random Forests \cite{herrmann2009website, armitage_iman} were shown to effectively classify traffic datasets, including encrypted TOR traffic. 
A survey of these classical models can be found in \cite{classicalMLsurveyraouf}. The use of time-series features including packet size and direction sequences, first suggested in \cite{herrmann2009website} as a side channel to attack TOR, was shown to be especially effective across models. 

With the widespread adoption of deep neural networks, numerous deep architectures were employed for ETC including Multi-Layer Perceptron, Convolutional Neural Networks, Long Short-Term Memory networks, Gated Recurrent Units, Stacked AutoEncoders, and Transformers (\cite{wang2017end, seq2img, FlowPic, fsnet, lotfollahi2020deep, Baseline, aceto2019mimetic, et-bert}) and shown to be effective on private datasets \cite{largescaleUCDavis, aceto2019mobile, Baseline} as well as open research datasets \cite{iscx-vpn, iscx-tor, quic-ucdavis}. By and large, some works such as \cite{seq2img, FlowPic, et-bert}  primarily focus on designing representations of traffic data for well-known architectures, while others like \cite{wang2017end, lotfollahi2020deep, Baseline, aceto2019mimetic,fsnet} focus on designing deep architectures tailored for well-known network traffic data representations. This paper focuses on automating the latter, more precisely, automatically designing an optimized architecture for a specific traffic data representation and a particular dataset. 

In an attempt to obtain a more realistic evaluation of the proposed ETC models in the literature, including classical and deep models, \cite{aceto2019mobile, rossilessonslearned} performed comparative studies of existing model performances on real-world datasets. \cite{aceto2019mobile} shows that the performances of deep and classical models are close, with 1D-CNN and MLP winning over Random Forest by a small margin. \cite{rossilessonslearned} highlights the importance of parameter tuning and shows that by simply tuning the maximum tree depth, XGBoost can significantly outperform CNN methods. 

The importance of hyper-parameter tuning is a driving factor in this work, although we take the opposite direction from \cite{rossilessonslearned}'s experiments with XGBoost, and focus on finding the best hyper-parameters and architecture for a CNN. Following our previous work, \cite{IFIPPaper}, in which we highlight the challenges along the way of manually adapting a successful deep architecture to new traffic traces, in this work, we explore a way to automate the manual process, given a dataset and a space of hyper-parameters.

AutoML was recently employed for traffic classification in \cite{nprint, ggfast}. Both \cite{nprint} and \cite{ggfast} focus on designing traffic representations for efficient classification performance. While \cite{nprint} proposes a normalized packet-level representation, \cite{ggfast} proposes a process for extracting flow-level features from sequences of packet size and directions. They both use the publicly available AutoGluon tool 
\cite{autogluon}. AutoGluon's approach to AutoML is different from NAS. Rather than opting to find the best model hyperparameters, it creates ensembles of several models by stacking them in layers. It is designed to be fast and simple to use by non-experts. Its base models include classical and deep models. For example, \cite{nprint} uses six types of base models including tree-based methods, deep neural networks, and neighbors-based classifiers. A similar ensemble-based method is employed in Mljar \cite{mljar}, a python AutoML package employed in \cite{automl_classical} for malware detection. However, Mljar only includes classical machine learning models as base models, hence, \cite{automl_classical} is further from our work than the former two. 
}

}

\section{AutoML4ETC} \label{sec:method}

In this section, we describe and discuss the major components of AutoML4ETC, namely: (i) the search space, (ii) the search algorithm, and (iii) the child model training strategy. These components are independent of one another, allowing flexibility in the choice of each component. 

\subsection{Search Space}

The search space in AutoML4ETC consists of a variety of operations (e.g., \textit{add}, \textit{concatenate}), connections (i.e., \textit{input}, \textit{output}) and their ordering, and the number of supported layers (e.g., convolution layers). These are the building blocks for the \textit{controller} (\ie \textit{search algorithm}) to choose from and interconnect to create the neural network architectures of child models. The output of any generated child model is connected to a \textit{Softmax} layer to produce the final classification.

As depicted in Figure~\ref{fig:automl_packetraw_design}, the overall architecture of a child neural network includes a \textit{single} Normal cell and a \textit{single} Reduction cell, which are composed of the  building blocks in the search space. In comparison to complex sequential architectures, this lightweight structure allows us to learn simpler and more generalizable models (\cf Section \ref{ss:automlresults}). The cell input goes through a hyper-layer before entering a cell, as shown in Figure \ref{fig:automl_packetraw_design}. The hyper-layer is either a \textit{Filter Alignment} or a \textit{Factorized Reduction} layer depending on the cell type.

\begin{figure}[tb]
\centering
\includegraphics[width=\columnwidth]{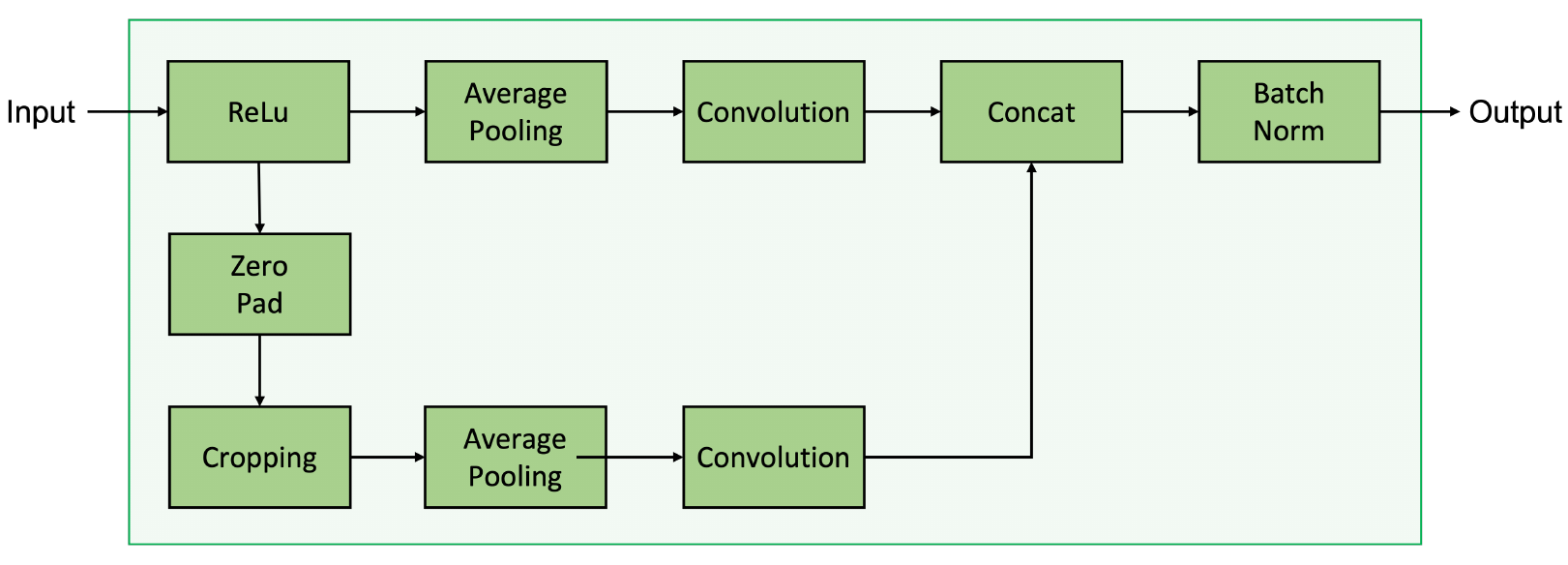}
\caption{Factorized Reduction hyper-layer modules}
\label{fig:factorized_reduction}

\end{figure}

\begin{itemize}
    \item \textit{Filter Alignment hyper-layer}: A Normal cell is preceded by a Filter Alignment layer, which consists of sequential ReLU, convolution, and batch normalization. This layer introduces filters at the beginning of the cell (\ie 64 initial filters, which is configurable).
    \item \textit{Factorized Reduction hyper-layer:} A Reduction Cell is preceded by a Factorized Reduction layer, depicted in Figure~\ref{fig:factorized_reduction}, which is necessary to process and reduce the input size in half. 
\end{itemize}

We compare the novel search space proposed for AutoML4ETC against state-of-the-art search spaces (\cf Section \ref{sec:evalsearchspace}).

\subsection{Search Algorithms}

\begin{figure}[tb]
\centering
\includegraphics[width=\columnwidth]{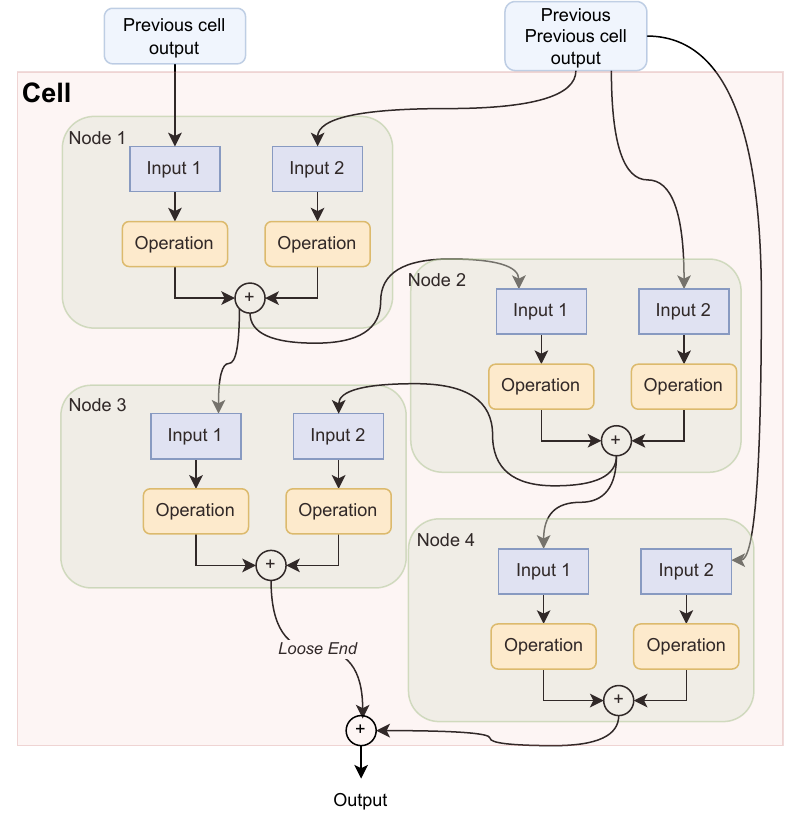}
\caption{Example cell components}
\label{fig:cell_inside}

\end{figure}

Given a search space, the search algorithm (\ie the controller) describes the child neural network architecture. To accomplish this, as discussed, the search algorithm composes a single Normal cell and a single Reduction cell from the search space. Each \textit{cell} consists of a number of  nodes (i.e., 4, which is empirically deduced and configurable), as shown in Figure~\ref{fig:cell_inside}. For each node, the search algorithm makes the following decisions:
\begin{enumerate}
    \item Choose \textit{Input 1} and \textit{Input 2} from the output of the previous nodes. If it is the first node, choose from the inputs of the \textit{cell}.
    \item Choose the operation for \textit{Input 1} and \textit{Input 2} from: (i) identity, (ii) separable convolution hyper-layer with kernel size 3 or 5, and (iii)
average or max pooling with kernel size 3. These choices are inspired from \cite{NAStoENAS}. 
    \item Add output of the two operations and return  as the output of the node.
\end{enumerate}

A \textit{separable convolution hyper-layer} consists of sequentially connected layers of ReLU, separable convolution, batch normalization, and dropout. 
We found that using a high-rate dropout layer (\eg 0.4) during search alleviates  overfitting to training data, leading to a more generalizable model. Furthermore, similar to \cite{NAStoENAS}, we employ two sequentially connected separable convolution hyper-layers every time the search algorithm chooses this operation. Such a structure for convolution layers has also been extensively used for deep ETC models \cite{Baseline, lotfollahi2020deep, largescaleUCDavis}, and has been shown to achieve high classification accuracy. 

Importantly, the search algorithm may not choose the output from a node as an input to other nodes in the same cell. We call these unused outputs \textit{Loose Ends}. Because Loose Ends may contain useful information 
, they are fed to the \textit{add} operation in the output of the last node of the cell.

Numerous state-of-the-art search algorithms have been proposed and evaluated in a setting with abundant resources (e.g., in excess of 400 GPUs \cite{EA, NAS, NAStoENAS}) and, or minimal to no time constraint (e.g., up to 50,000 trials \cite{NAStoENAS}, 310 epochs for architecture search \cite{ENAS}). Indeed, in a resource- or time-constrained environment, these algorithms may not converge to their optimal performance. Moreover, simple Random Search (RS) has proven very competitive  against its more complex counterparts. 
In \cite{MCTS}, the authors compared a variation of EA and MCTS to RS with the spectrum of measured accuracy ranging between 94.1\% and 94.2\%. Similarly, authors in \cite{NAS, ENAS, EA} showed RS performance to be within 1\% of other search algorithms.

The modularity of AutoML4ETC allows to choose any search algorithm independently from the search space. Therefore, we investigate different state-of-the-art neural architecture search algorithms, and evaluate their impact on the performance and complexity of the best child model. Specifically, in AutoML4ETC, we experiment with RL \cite{ENAS}, MCTS \cite{MCTS}, and EA \cite{EA}, and compare them against the baseline RS algorithm. With the exception of RS, these search algorithms were developed and evaluated on the NASNet or ENAS Micro search spaces, which were an inspiration for our novel search space in AutoML4ETC. Therefore, comparing these searching algorithms in AutoML4ETC is particularly relevant (\cf Section \ref{sec:seachalgocomp}). 

\subsection{Child Model Training Strategies}

NAS is inherently time consuming and includes the following steps: (i) search algorithm, (ii) child architecture composition, and (iii) training of child models. In our experiments, we estimated the time taken for the first two steps to be around a 100 milliseconds, which is negligible in comparison to the time in training child models. Indeed, every time a child architecture is composed, it is trained over several epochs and then evaluated for performance. The performance is a reward signal for a child model, and  helps the search algorithm to converge to better child neural networks in future trials. Our goals is to reduce the child model training time (\ie number of epochs) without compromising NAS performance.

We investigate two types of child model training strategies in AutoML4ETC, namely \emph{full training} and \emph{partial training}. Full training pertains to training child models on as many epochs as needed to generate the best possible architecture. In contrast, in partial training, the child models are trained on a smaller number of epochs, and only the best-performing child model is trained for additional epochs, as needed. 

For example, the child models in the partial training strategy could be trained for 25\% of the full training epochs, with the best performing child model trained further for the remaining 75\% epochs. This could significantly reduce NAS time. However, the best child model from partial training is only an estimate of the global best child model. Therefore, there is an interesting trade-off to consider between the two training strategies in AutoML4ETC (\cf Section \ref{sec:trainstrategycomp}).

\begin{figure*}[th]
\centerline{
\includegraphics[width=\textwidth]{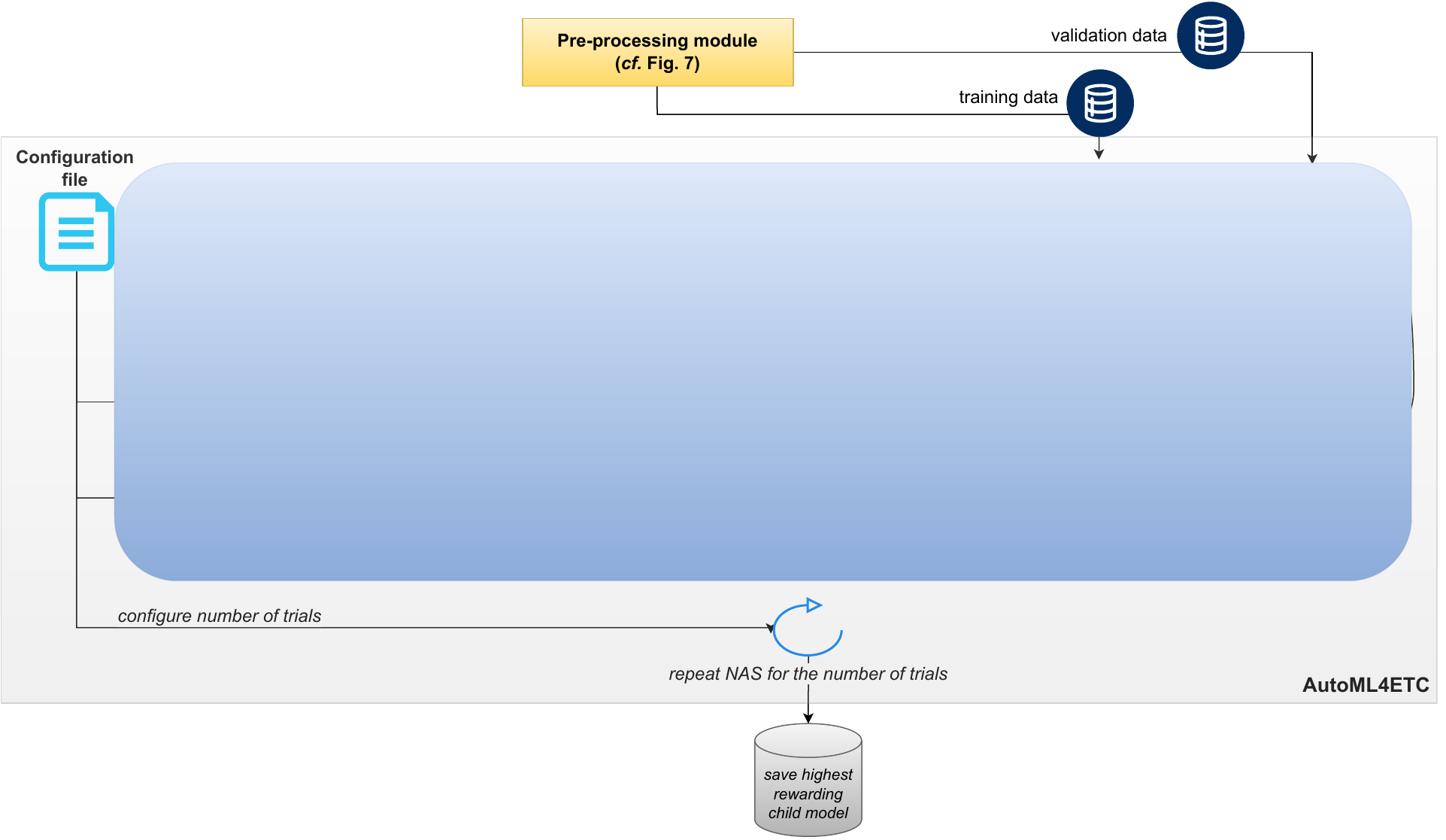}}
\caption{Overall AutoML4ETC framework}
\label{fig:AutoML4ETC_HighLevel}

\end{figure*}

\section{Evaluation} \label{sec:eval}

We start by presenting the overall settings and methodology for evaluating AutoML4ETC. \blue{The high-level workflow of the AutoML4ETC framework, including all of its components, is depicted in Fig. \ref{fig:AutoML4ETC_HighLevel}.} In Section \ref{s:baseline}, we describe the baseline state-of-the-art ETC models used for comparing the performance of the AutoML4ETC-generated models. In Section~\ref{ss:metrics}, we present the performance criteria and metrics used in our evaluation. We describe the pre-processing and labeling process of TLS and QUIC raw packet captures (\ie PCAP files) as well as the datasets used in our experiments in Section~\ref{s:dataset}.

Our evaluation study is four-fold. We evaluate the search space of AutoML4ETC and compare it against a baseline search space in Section~\ref{sec:evalsearchspace}. We evaluate the performance of different search algorithms and child-model training strategies in Section~\ref{sec:seachalgocomp} and Section~\ref{sec:trainstrategycomp}, respectively. Finally, in Section~\ref{ss:automlresults}, we compare the performance of AutoML4ETC-generated architectures against the baseline state-of-the-art ETC models on various datasets.

\subsection{State-of-the-art ETC Models}\label{s:baseline}
We use four baseline state-of-the-art ETC models that have shown superior performance on the input features employed in this work. 

The \textit{UWOrange} \cite{Baseline} model consists of three different branches each tailored to a specific type of input feature type. 
The outputs of the three parts are concatenated and further processed by dense layers followed by a Softmax layer. One branch of the model consists of convolutional layers that process \textit{packet raw bytes}. We call this branch the \textit{UWOrange-H} model\blue{, and use it as a baseline for comparison}.

The \textit{UCDavis CNN} \cite{largescaleUCDavis} model is designed to operate on \textit{packet raw bytes}. It differs from the UWOrange-H model in terms of the number of units in the last dense layers, and it does not use dropout layers. However, the essential part of the UCDavis CNN and the UWOrange-H models, \ie the CNN layers, are similar. Another difference between the two models is in their input. The raw packet data in UCDavis CNN includes the first six packets of a flow, while the input data in UWOrange-H consists of the first 600 bytes of the first three TLS handshake packets of the flow.

\blue{The \textit{DeepPacket CNN} \cite{lotfollahi2020deep} is a high-performing DL model designed specifically for \textit{packet raw bytes}. This CNN model receives a 1,500-dimensional vector as input and comprises two 1D convolutional layers placed one after the other, followed by a max-pooling layer. The vector is flattened and fed through four fully connected layers, with the final layer serving as the Softmax classifier.

\textit{E2E CNN} \cite{E2E} is the first encrypted traffic classifier that employs 1D-CNNs. It was shown to outperform traditional ML methods. This model inspired many other CNN architectures including all the state-of-the-art baseline models considered in this paper. Furthermore, this CNN-based model is also tailored for \textit{packet raw bytes}, making it a valid baseline to compare the performance of AutoML4ETC. E2E CNN receives a 768-dimensional vector as input and comprises one 1D convolutional layer, followed by a max-pooling layer. The same structure is repeated and finally, the vector is flattened and fed to two fully-connected layers with the final layer serving as the Softmax classifier.

The detailed model architectures implemented in our experiments are depicted in Appendix Fig.~\ref{fig:sota}.}

\subsection{Metrics and Settings}\label{ss:metrics}

A binary classification task may result in True Positives (TP), False Positives (FP), True Negatives (TN), and False Negatives (FN). The \textit{precision}, \textit{recall}, \textit{F1-score} and \textit{accuracy} of such a classifier are derived as follows:

\begin{equation*}
\begin{split}
    Precision & = \frac{TP}{TP+FP} \times 100 \\
    Recall & =\frac{TP}{TP+FN} \times 100 \\
    F1-score & =\frac{2 \times Precision \times Recall}{Precision+Recall} \times 100\\
    Accuracy & = \frac{TP+TN}{TP+TN+FP+FN} \times 100\\
\end{split}
\end{equation*}

\vspace{5pt}
In this work, we aim to use AutoML4ETC to generate multi-label encrypted network traffic classifiers. As such, we evaluate the performance of a generated model in terms of its \textit{weighted average recall}, \textit{weighted average precision}, and \textit{weighted average F1-score}, by averaging the recall, precision, and F1-score of the classifier across all classes weighted by the fraction of flows belonging to each class in the dataset. We also evaluate the \textit{accuracy} of the model as the fraction of correctly classified flows. 

Each child network is trained with an initial learning rate of 0.001. Furthermore, while the child models are being trained, we cut the learning rate in half every 10 epochs. We found this to be a more effective solution than a fixed learning rate, as it increases the resolution of the gradient descent's search as the search progresses. Moreover, we use the \textit{Adam} optimizer and the sparse categorical cross-entropy loss function to train the child networks. Further details can be found in the source code (cf. Footnote \ref{{foot:code}}).

\blue{Our software stack for neural architecture design and pre-processing consists of Tensorflow 2.2~\cite{tensorflow2015-whitepaper} with Keras~\cite{chollet2015keras} and PySpark 2.4.4~\cite{spark}, and a custom version of Hypernets 0.2.3~\cite{Hypernets} with custom HyperKeras.}

\subsection{Datasets and Pre-processing}\label{s:dataset}

\begin{figure}[tb]
\centering
\includegraphics[width=\columnwidth]{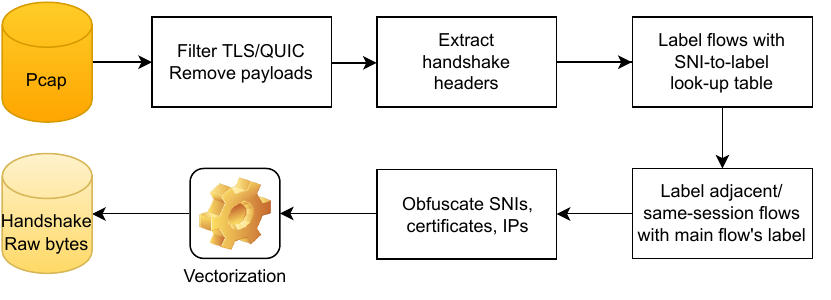}
\caption{Pre-processing procedure for real-world datasets}
\label{fig:preproc}

\end{figure}

 The overall procedure for pre-processing and labeling raw packet captures (\ie PCAP files) into ML-usable datasets is shown in Figure \ref{fig:preproc}. The first pre-processing step consists of removing the packet payloads beyond the TLS or QUIC header. The resulting PCAP files are broken into \textit{flows}, where each flow consists of packets close in time that share source IP, destination IP, source port, destination port, and protocol. From each flow, the headers of the first three TLS handshake packets are extracted. The TLS Server Name Indication (SNI) header is used for labeling the flow and is obfuscated next, and so is the TLS cipher information header field. The IP addresses are also masked. For the QUIC flows, only the first \textit{ClientHello} packet in the handshake phase is extracted. The SNI header is used for labeling the flow and is obfuscated next, and so is the TLS cipher information header. We do not include more than one packet for QUIC flows as the latter packets are encrypted and do not have any information useful to the classifier.
 
 The real-world datasets were labeled based on the SNI field in each flow, which is one reason why we obfuscate the SNI value in preprocessing. Not all flows contain a readable SNI value. Moreover, the utilization of clear SNI is likely to decline in favor of the proposed Encrypted SNI (ESNI) extension \cite{ietf-tls-esni-14}. Hence, we need traffic classifiers that learn the intrinsic characteristics of the flows rather than a mapping from SNI to traffic classes.  

We used an approximate labeling function to extract labels from SNIs. We developed a look-up table by visiting top websites in each service class and extracting regular expressions from their domain names, which are matched with the SNI value to map each SNI to a label. Because not all flows contain an SNI value, we also label adjacent flows (\ie flows with the same TLS session-id or close-enough start time) based on the main flow that has an SNI to increase the number of labeled flows.

The labeling module was initially designed for the TLS datasets and then adapted to the QUIC dataset. We observed from the SNIs in the QUIC dataset that some of the TLS dataset classes did not have any instances in the QUIC dataset. Similarly, some of the classes are unique to QUIC, such as \textit{resources} and \textit{e-commerce}.

The \textit{Resources} class contains flows that add material to the webpage, such as Javascript, CSS, and HTML libraries, and often have one end in some major hosting service, \eg Cloudflare. On the other hand, flows that belong to third-party Advertisement services are categorized under \textit{ecommerce}.

We categorize QUIC and TLS using different sets of classes for two reasons: \textit{(i)} QUIC is still a relatively new protocol and is not yet as widely adopted as TLS, and \textit{(ii)} QUIC offers a higher connection speed than HTTP over TLS; therefore, time-sensitive services such as \textit{resources} adopt it to enhance the loading time of their clients' websites. However, it is less frequently adopted by services such as mail or file download, where loading time is not crucial.

\begin{table*}[tb!]
\caption{Service-level datasets properties}
\label{tab:dataset-properties}
\centering
\resizebox{2\columnwidth}{!}{%
\begin{tabular}{|c|c|c|c|c|}
\hline
Dataset Type &
  Dataset name &
  \begin{tabular}[c]{@{}c@{}}Number of \\ flows (thousand)\end{tabular} &
  \begin{tabular}[c]{@{}c@{}}Number of labeled\\ flows (thousand)\end{tabular} &
  Dataset classes \\ \hline\hline
\multirow{5}{*}{\begin{tabular}[c]{@{}c@{}}Real-world (Orange)\\ Protocol: TLS\end{tabular}} &
  July 2019 &
  762.7 &
  119.8 &
  \multirow{5}{*}{\begin{tabular}[c]{@{}c@{}}chat, download, games,\\ mail, search, social,\\ streaming, Web\end{tabular}} \\ \cline{2-4}
 &
  September 2020 &
  411.7 &
  89.9 &
   \\ \cline{2-4}
 &
  April 2021 &
  284.8 &
  42.3 &
   \\ \cline{2-4}
 &
  May 2021 &
  124.0 &
  51.2 &
   \\ \cline{2-4}
 &
  June 2021 &
  261.2 &
  51.2 &
   \\ \hline
\begin{tabular}[c]{@{}c@{}}Real-world (Orange)\\ Protocol: QUIC\end{tabular} &
  QUIC - May 2021 &
  37.8 &
  26.0 &
  \begin{tabular}[c]{@{}c@{}}web, social, streaming,\\ ecommerce, resources, games\end{tabular} \\ \hline
\begin{tabular}[c]{@{}c@{}}Semi real-world (Controlled environment)\\ Protocol: Mixed\end{tabular} &
  ISCXVPN2016\cite{icxdataset} &
  140.7 &
  140.7 &
  \begin{tabular}[c]{@{}c@{}}chat, email, file transfer, streaming, torrent, voip\\ VPNchat, VPNemail, VPNfile-transfer,\\ VPNstreaming, VPNtorrent, VPNvoip\end{tabular} \\ \hline
\end{tabular}%
}

\end{table*}

\begin{table*}[tb!]
\caption{Application-level datasets properties}
\label{tab:dataset-app}
\begin{center}
\resizebox{2\columnwidth}{!}{%
\begin{tabular}{|c|c|c|c|c|}
\hline
Dataset Type &
  Dataset name &
  \begin{tabular}[c]{@{}c@{}}Number of\\ flows (thousand)\end{tabular} &
  \begin{tabular}[c]{@{}c@{}}Number of labeled\\ flows (thousand)\end{tabular} &
  Dataset classes \\ \hline\hline
\multirow{9}{*}{\begin{tabular}[c]{@{}c@{}}Real-world  (Orange)\\ Protocol: TLS\end{tabular}} &
  \multirow{2}{*}{July 2019} &
  \multirow{2}{*}{762.7} &
  \multirow{2}{*}{83.1} &
  \multirow{9}{*}{\begin{tabular}[c]{@{}c@{}}chatFacebook, chatSnapchat, \\ chatWhatsApp, downloadApple,\\ downloadGooglePlay, mailGmail,\\ mailHotmail, mailOutlook, searchGoogle,\\ socialFacebook, socialInstagram, socialTwitter,\\ streamingFacebook, streamingNetflix,\\ streamingSnapchat, streamingYoutube,\\ webAmazon, webAppleLocalication, \\ webMicrosoft
  \end{tabular}} \\
 &
   &
   &
   &
   \\ \cline{2-4}
 &
  \multirow{2}{*}{September 2020} &
  \multirow{2}{*}{411.7} &
  \multirow{2}{*}{59.8} &
   \\
 &
   &
   &
   &
   \\ \cline{2-4}
 &
  \multirow{2}{*}{April 2021} &
  \multirow{2}{*}{284.8} &
  \multirow{2}{*}{26.3} &
   \\
 &
   &
   &
   &
   \\ \cline{2-4}
 &
  \multirow{2}{*}{May 2021} &
  \multirow{2}{*}{124.0} &
  \multirow{2}{*}{11.1} &
   \\
 &
   &
   &
   &
   \\ \cline{2-4}
 &
   June 2021 &
   261.2 &
   34.0 &
   \\ \hline
\multirow{3}{*}{\begin{tabular}[c]{@{}c@{}}Real-world (Orange)\\ Protocol: QUIC\end{tabular}} &
  \multirow{3}{*}{QUIC - May 2021} &
  \multirow{3}{*}{37.8} &
  \multirow{3}{*}{9.3} &
  \multirow{3}{*}{\begin{tabular}[c]{@{}c@{}}chatDiscord, resourcesGoogle, resourcesMgid, \\ resourcesPbstck, resourcesSmpush,\\ resourcesJSDelivr,socialFacebook, webCanva\end{tabular}} \\
 &
   &
   &
   &
   \\
 &
   &
   &
   &
   \\ \hline
\begin{tabular}[c]{@{}c@{}}Synthetic (UCDavis)\\ Protocol: QUIC\end{tabular} &
  UCDavis QUIC\cite{quic-ucdavis} &
  3.63 &
  3.63 &
  \begin{tabular}[c]{@{}c@{}}GoogleDoc, GoogelDrive, GoogleMusic,\\ GoogleSearch, YouTube\end{tabular} \\ \hline
\begin{tabular}[c]{@{}c@{}}Semi real-world (Controlled environment)\\ Protocol: Mixed\end{tabular} &
  ISCXVPN2016\cite{icxdataset} &
  41.7 &
  41.7 &
  \begin{tabular}[c]{@{}c@{}}AIM chat, email, facebook, FTPS, scp\\ gmail, hangouts, ICQ, netflix, sftp, youtube\\ skype, spotify, torrent, tor, vimeo, voipbuster\end{tabular} \\ \hline
\end{tabular}%
}
\end{center}

\end{table*}

We evaluate AutoML4ETC on six real-world datasets and three public benchmark datasets of network traffic generated either synthetically or in controlled environments \cite{quic-ucdavis, icxdataset}. Five of the real-world datasets consist of TLS flows and the sixth consists of QUIC flows. The real-world datasets were generated by pre-processing and labeling traffic traces captured on the Orange mobile network, and named after their year and month of capture. Table~\ref{tab:dataset-properties} highlights the properties of the datasets labeled at the service level. We also experiment with application-level classification. \blue{The application-level labels are listed in Table~\ref{tab:dataset-app} and differ from the service-level labels}. In each dataset, we reserve 80\% of the labeled data samples for training and the remaining 20\% for evaluation. \blue{For the real-world datasets we use a 600-byte cut-off as suggested in~\cite{Baseline}. For the other datasets, we use the pre-processed version made available by the original authors.}

\subsection{Evaluation of the search space}\label{sec:evalsearchspace}

In order to show how the AutoML4ETC search space surpasses the performance of current state-of-the-art ETC, we build another search space based on the state-of-the-art ETC architectures, which we name \textit{CNN + MLP}. We then compare the performance of the best model found in this search space with that of the best model found in AutoML4ETC. The best models are found using the same RS algorithm.

\begin{table*}[tb]
\caption{CNN search space parameters, values, and types}
\label{tab:cnn_params} 
\centering
{\fontsize{7}{5}\selectfont
\begin{tabular}{|c||c|c|c|c|c|c|c|c|}
\hline
Parameter &
  CONV Block Repeat &
  Kernel size &
  Filter size &
  Dropout &
  Order &
  Pooling layer &
  Activation &
  Batch norm \\ \hline\hline
Values &
  {[}2, 3, 4, 5, 6{]} &
  {[}(1, 1), (2, 2){]} &
  {[}32, 64{]} &
  (0, 0.05) &
  {[}dropout, activation, batch norm{]} &
  {[}MaxPool, AveragePool{]} &
  {[}relu, elu{]} &
  Boolean \\ \hline
Type &
  Choice &
  Choice &
  Choice &
  Real &
  Permutation &
  Choice &
  Choice &
  Optional \\ \hline
\end{tabular}%
}

\end{table*}

\begin{table*}[tb]
\caption{MLP search space parameters, values, and types}

\label{tab:MLP-SS}
\centering
{\fontsize{7.5}{5}\selectfont
\begin{tabular}{|c||c|c|c|c|c|c|c|}
\hline
Parameter & Dense units             & Number of Dense layers  & Reduce factor     & Activation      & Batch norm & Dropout     & Order                                     \\ \hline\hline
Values    & {[}100, 200, 400{]} & {[}3, 4, 5{]} & {[}1, 0.7{]} & {[}relu, elu{]} & Boolean    & (0.3, 0.5)     & {[}dropout, batch norm, activation{]} \\ \hline
Type      & Choice                  & Choice        & Choice            & Choice          & Optional   & Real number & Permutation                           \\ \hline
\end{tabular}}

\end{table*}

\subsubsection*{\textbf{CNN + MLP}}
\begin{figure}[tb]
\centering
\includegraphics[width=\columnwidth]{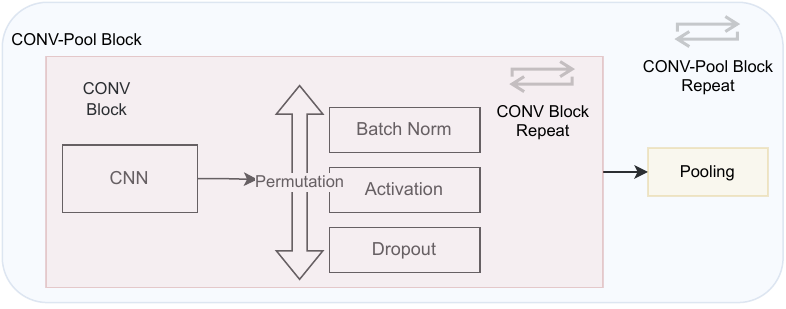}
\caption{Overview of CNN search space}
\label{fig:cnn-ss}
\end{figure}

The \textit{CNN + MLP}  search spaces are inspired by the ETC state of the art \cite{seq2img, Baseline, largescaleUCDavis, quic-ucdavis, lotfollahi2020deep} and covers their architecture. The overall structure of the CNN (\ie 2D CNNs or 1D CNNs) search space is depicted in Figure~\ref{fig:cnn-ss}. Each \textit{CONV-Pool Block} is a sequence of one or more \textit{CONV Blocks} connected to a pooling layer. The number of CONV-Pool Blocks in the sequence is determined by the \textit{CONV-Pool Block Repeat} parameter. Moreover, the inner CONV Block is also sequentially repeated  \textit{CONV Block Repeat} number of times, where the two repeat parameters are independent of one another. Additionally, for the first two CONV-Pool Blocks, we set the CONV Block Repeat to two repetitions. Thereafter, CONV Block Repeat can be from a range of 3 to 5, which is a number of convolutional layers often used in ETC \cite{seq2img, Baseline, largescaleUCDavis, quic-ucdavis, lotfollahi2020deep}. Also, the number of filters is cut in half in the following repetitions of CONV Block, \ie after the first two CONV-Pool Blocks. The parameters for this search space are summarized in Table~\ref{tab:cnn_params}. \textit{Choice} means that the search algorithm chooses a value from a range of \textit{values} for that parameter; \textit{Optional} means that this layer is optional; \textit{Permutation} only organizes the best permutation of values; \textit{Reduce factor} defines the sequential amount of reduction in dense units.

\begin{figure}[tb]
\centering
\includegraphics[width=\columnwidth]{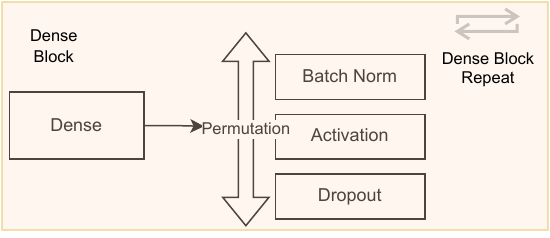}
\caption{Overview of MLP search space}
\label{fig:mlp_ss}
\end{figure}

We connect \textit{CNN} search space to the \textit{MLP} search space to construct the \textit{CNN + MLP} search space. For the MLP search space, we use a sequence of dense layers and a permutation of dropout, batch norm, and activation layers between every two dense layers, as shown in Figure~\ref{fig:mlp_ss}. We repeat this \textit{Dense Block} several times, which is specified by the \textit{Dense Block Repeat} parameter. These blocks are connected and the number of units in the dense layer is reduced in the next block by the \textit{Reduce factor} parameter. The parameters of the MLP search space are summarized in Table \ref{tab:MLP-SS}.

\subsubsection*{\textbf{Performance Comparison}} 

We showcase the performance of AutoML4ETC against CNN + MLP by comparing the performances of the best model found in each search space. The same RS algorithm is used to search all search spaces, as our goal is to evaluate the search space, not the search algorithm. Table~\ref{tab:searchspace_compare_rawbytes} shows the best model's performance on the \textit{May 2021} TLS dataset, when the child model is trained for 40 epochs. We use the May 2021 TLS dataset as it is the smallest of our TLS datasets for timing purposes. We will justify the 40 epochs and report results with different numbers of epochs and other datasets in later sections.

The best child model generated from the \textit{CNN-2D + MLP} search space after 200 trials, achieves a 77.55\% accuracy and has almost 22 million parameters. The input to the models in this search space is transformed into two-dimensional images

(\ie \cite{seq2img} with NAS). 
However, turning raw bytes into 1D vectors and using 1D CNNs can boost the performance of the model. We speculate that the arbitrary arrangement of sequential packet bytes in 2D possibly confuses the model, as it considers learning patterns between bytes that are put next to one another in 2D for no reason.
Evidently, on the \textit{CNN-1D + MLP} search space, it is possible to obtain a child model that achieves 78.75\% accuracy with almost half the number of parameters after the same number of trials.

RS on the \textit{ENAS Micro} search space can find a much lighter model (\ie 120.6 thousand parameters) that achieves higher accuracy (\ie 80.4\%), after 200 trials as well. However, with the \textit{AutoML4ETC} search space, it is possible to generate a classifier that is not only more accurate (\ie 82.86\% classification accuracy) but also lighter (\ie 111.5 thousand parameters) than any of the above classifiers after only half of those trials (\ie 100 trials). Hence our \textit{AutoML4ETC} search space outperforms state-of-the-art search spaces both in terms of accuracy and complexity of the best child model on this dataset.

\begin{table}[tb]
\centering
\footnotesize
\caption{Packet raw bytes search spaces comparison}
\label{tab:searchspace_compare_rawbytes}
\begin{tabular}{|c|c|c|c|c|}
\hline
\begin{tabular}[c]{@{}c@{}}Search\\ Space\end{tabular} &
  \begin{tabular}[c]{@{}c@{}}Search\\  Algorithm\end{tabular} &
  Trials &
  \begin{tabular}[c]{@{}c@{}}Accuracy\\ (\%)\end{tabular} &
  \begin{tabular}[c]{@{}c@{}}Parameters\\ (Thousand)\end{tabular} \\ \hline\hline
\textbf{AutoML4ETC} & \textbf{RS} & \textbf{100} & \textbf{82.86} & \textbf{111.5} \\ \hline
2D-CNN + MLP        & RS          & 200          & 77.55          & 21,940.5       \\ \hline
1D-CNN + MLP        & RS          & 200          & 78.75          & 12,116.1     \\ \hline
ENAS Micro          & RS          & 200          & 80.4           & 120.6          \\ \hline
\end{tabular}%

\end{table}

\subsection{Comparison of Search Algorithms}\label{sec:seachalgocomp}
We discuss and compare the performances of different search algorithms in this section after briefly describing the algorithms. 
\subsubsection*{\textbf{RL}}
Based on the RNN controller in \cite{ENAS}, which uses \textit{REINFORCE} with baseline and \textit{Adam} as the optimizer. The RNN controller is a single-layer Long Short-term Memory (LSTM) with 100 units applied to the controller logits. Baseline decay is set to 0.999, while the norm of gradients is clipped at 5.0. The learning rate for the \textit{Adam} optimizer is 1e-3.

\subsubsection*{\textbf{MCTS}}
Based on \cite{MCTS} and uses \textit{UCT} \cite{UCT}. The maximum node expansion is set to 10. When rolling out, the sample size for the simulation network to assess potential pathways is set to 10. The simulation network is \textit{LightGBM} \cite{Ke2017LightGBMAH} with default parameters.

\subsubsection*{\textbf{EA}}
Based on \cite{EA}, where the population size is set to 20. The number of parent candidates selected per evolution cycle is set to 5. 

\subsubsection*{\textbf{RS}}
Makes random decisions at each step. It does not maintain a state to update itself based on previous decisions.

For a fair comparison, we fix the other parameters of  AutoML4ETC as follows. We set the number of child model training epochs to 10 for faster training and set the total number of trials to 100. We will discuss the effectiveness of the 10 epochs of child model training in the next section.

We compare the mean accuracy of the top-N child models found by each of the search algorithms for N=1, 5, 10, 20, and 30. This is because we are not only interested in comparing the performance of the global best child models, but also we want to compare the overall ability of each search algorithm to find reasonably good (\ie reasonably accurate) child models throughout the search process. As an example, top-5 child models are the 5 child models that achieve the best validation accuracy sorted in descending order.

\begin{figure}[tp]
\centering
\includegraphics[width=0.9\columnwidth, height=0.6\columnwidth]{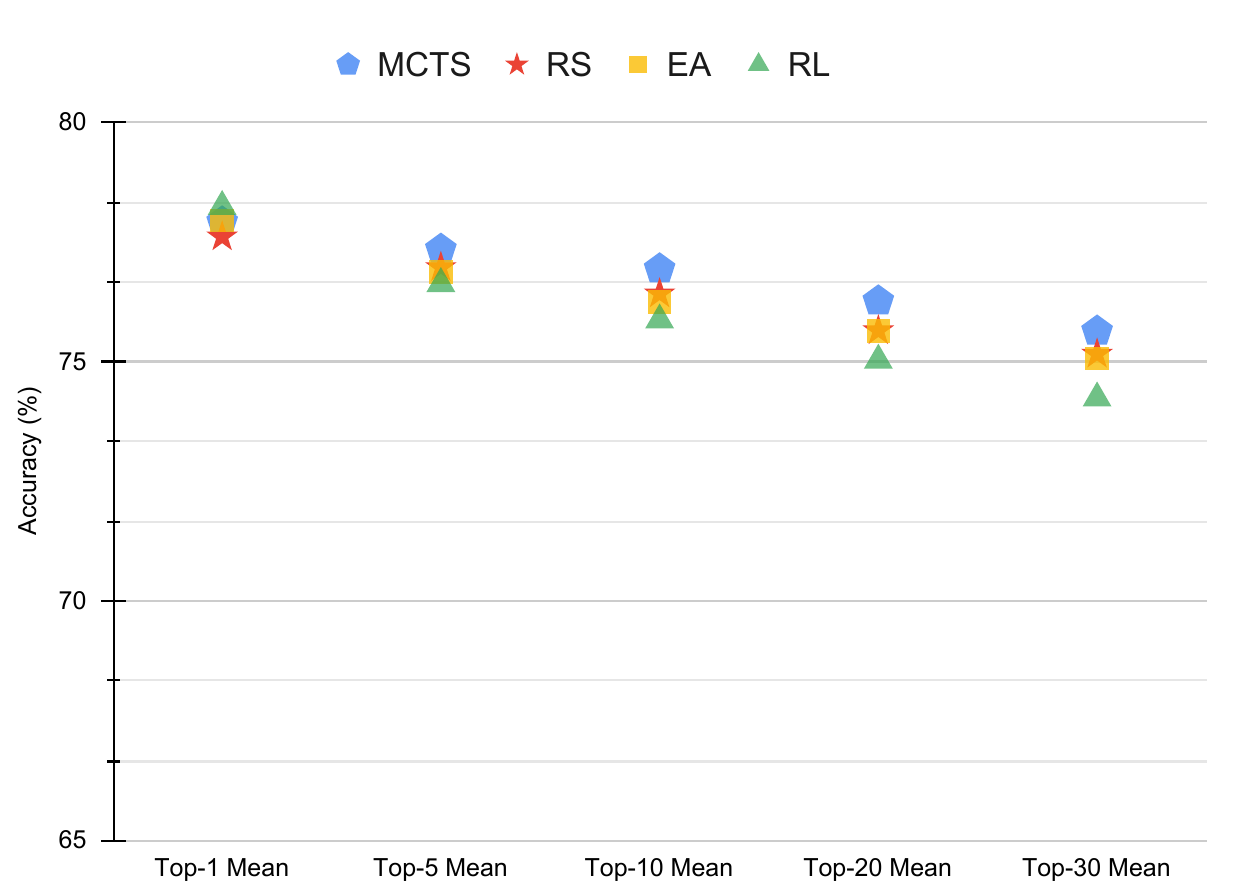}
\caption{Comparison of average accuracy of top-N child models using different search algorithms}
\label{fig:search_algorithms}

\end{figure}

Performance evaluation results are depicted in Figure~\ref{fig:search_algorithms}. Several interesting observations can be made here. First, concerning the global best child models (\ie top-1 child models), it is evident that all of the search algorithms lead to equally well-performing best child models; the standard deviation of the accuracy distribution across the best child models is only 0.25\%. This result highlights, in particular, the power of our search space, where the simplest search algorithm (\ie RS) can perform as well as much more complex search algorithms at the cost of a few more parameters (\ie 206.53, 231.62, 242.12, and 258.24 thousand for RL, MCTS, EA, and RS, respectively).

Another interesting observation is that as N increases, the gap between the mean accuracies of the top-N child models grows. While the MCTS algorithm is the top performer for N=5,..,30, RL performs the worst. Moreover, in this same range, the EA and RS algorithms perform almost identically and score in between MCTS and RL. This suggests that MCTS can build more top-performing models with fewer trials and child model training epochs. 
 
 All of these findings and observations suggest that designing a good search space is more important than the search algorithm itself when using fewer trials and epochs for child model training. A good search space is when most of the architecture combinations result in reasonably good accuracy, and this applies to our search space. Therefore, in the state of the art versus AutoML4ETC section, for the sake of efficiency in the number of parameters with the best accuracy, we choose to use RL as our search algorithm. RL showed the best performance for the top-1 child model and also found the best child model with the least number of parameters. However, the search spaces in AutoML4ETC can be combined with another search algorithm to realize any other desiderata.

\subsection{Evaluation of Child Model Training Strategies}\label{sec:trainstrategycomp}

\begin{figure}[tb!]
\centering
\includegraphics[width=0.9\columnwidth]{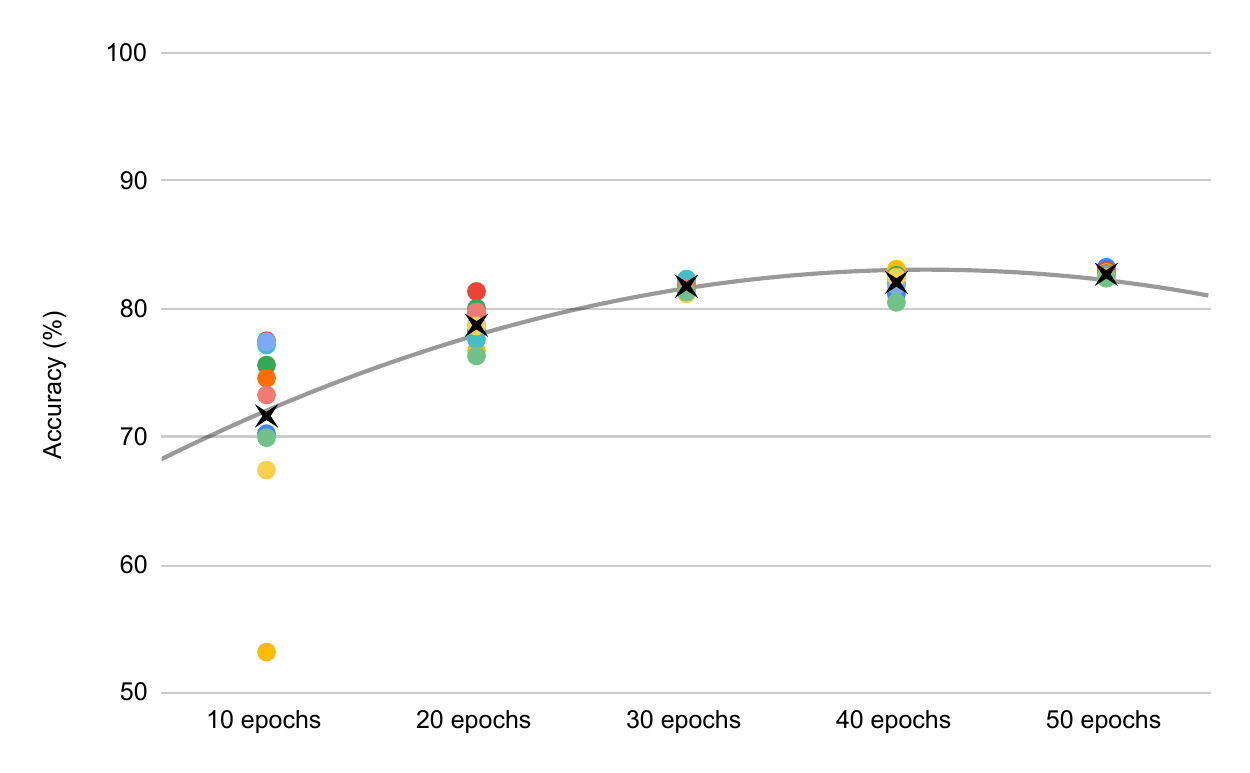}
\caption{Accuracy of top-10 child models with different training epochs; the $\times$ mark is the average for each epoch.}
\label{fig:flattening_H_side}

\end{figure}

We start by searching for an upper bound for the number of training epochs needed to find the best child model. To find this number, we conduct a set of experiments where we increase the number of training epochs starting from 10 and validate every 10 epochs on the testing dataset. Figure~\ref{fig:flattening_H_side} is the mean validation accuracy of the top-10 child models for a varying number of training epochs. We can see that the average accuracy of the top-10 child models (\ie black line curve) flattens beyond 40 epochs. Therefore, 40 epochs can be set as our upper bound for the full training of child models.

From another perspective, if we partially train the child models over 10 epochs only, we can save approximately 75\% of the total NAS time. With this method we just use 10 epochs for training child models, then extract the top child model and train it for an extra 30 epochs. However, the best child model resulting from partial training is an estimate of the global best child model.

Table~\ref{tab:partialVSfull_rawbytes} further shows the trade-off between the accuracy and complexity of the top child model. We can see that with a 10-epoch partial training strategy, the top child model achieves 79.71\% accuracy. However, with the full training strategy, \ie training over 40 epochs, the top child model can achieve a higher accuracy of 82.86\%. Additionally, the number of parameters of the top child model resulting from the full training approach is less than half of its counterpart. Therefore, the trade-off can be diluted down to a loss of $\sim$3\% in accuracy with twice as many parameters for a $\sim$75\% lower time complexity (\ie search time). 

Since the AutoML4ETC models are light in general with both strategies (\ie number of parameters) and the difference in accuracy is noticeable, a 40-epoch full training strategy seems to be a better option for this particular search space.

\begin{table}[tb]
\caption{Performance of partial training of child models (i.e., 10 epochs) vs. full training strategy (i.e., 40 epochs)}
\label{tab:partialVSfull_rawbytes}
\centering
\footnotesize
\begin{tabular}{|c|c|c|c|}
\hline
\begin{tabular}[c]{@{}c@{}}Child model\\  training epochs\end{tabular} &
  \begin{tabular}[c]{@{}c@{}}Child model\\ accuracy (\%)\end{tabular} &
  \begin{tabular}[c]{@{}c@{}}Full train\\  accuracy (\%)\end{tabular} &
  \begin{tabular}[c]{@{}c@{}}Total \\ parameters\end{tabular} \\ \hline\hline
10 epochs &
  77.61 &
  79.72 &
  263,368 \\ \hline
40 epochs &
  82.86 &
  - &
  111,560 \\ \hline
\end{tabular}%
\end{table}

\begin{table*}[tb!]
\caption{AutoML4ETC versus the state-of-the-art models in service-level classification}
\label{tab:automl_vs_baselines_rawbytes}
\centering
\scriptsize
\begin{tabular}{|c|c|c|c|c|c|c|c|}
\hline
Dataset &
  Model &
  \begin{tabular}[c]{@{}c@{}}Accuracy\\ (\%)\end{tabular} &
  \begin{tabular}[c]{@{}c@{}}W Avg.\\  F-1 score (\%)\end{tabular} &
  \begin{tabular}[c]{@{}c@{}}W Avg. \\ recall (\%)\end{tabular} &
  \begin{tabular}[c]{@{}c@{}}W Avg. \\ precision (\%)\end{tabular} &
  Total parameters &
  Trainable parameters \\ \hline \hline
\multirow{4}{*}{July 2019}       & \textbf{AutoML4ETC} & \textbf{95.86} & \textbf{95.86} & \textbf{95.86} & \textbf{95.87} & \textbf{271,560} & \textbf{266,696} \\ \cline{2-8} 
                                 & UWOrange-H              & 94.87          & 94.87          & 94.87          & 94.93          & 7,588,360        & 7,588,360        \\ \cline{2-8} 
                                 & UCDavis CNN        & 90.95          & 90.92          & 90.95          & 90.93          & 6,507,016        & 6,507,016        \\ \cline{2-8} 
                                 & DeepPacket CNN      & 90.49          & 90.47          & 90.49          & 90.48          & 9,960,732        & 9,960,732       
                                 \\ \cline{2-8} 
                                 & E2E CNN         & 87.40          & 87.39          & 87.40          & 87.46          & 11,202,440        & 11,202,440       
                                 \\ \hline
\multirow{4}{*}{September 2021}  & \textbf{AutoML4ETC} & \textbf{87.04} & \textbf{87.97} & \textbf{87.04} & \textbf{91.09} & \textbf{156,104} & \textbf{153,032} \\ \cline{2-8} 
                                 & UWOrange-H              & 83.59          & 84.48          & 83.59          & 87.84          & 7,588,360        & 7,588,360        \\ \cline{2-8} 
                                 & UCDavis CNN        & 86.52          & 86.49          & 86.52          & 86.52          & 6,507,016        & 6,507,016        \\ \cline{2-8} 
                                 & DeepPacket CNN      & 87.38          & 87.39          & 87.38          & 87.42          & 9,960,732        & 9,960,732       
                                 \\ \cline{2-8} 
                                 & E2E CNN          & 81.46          & 81.59         & 81.46          & 81.98         & 11,202,440       & 11,202,440        
                                 \\ \hline
\multirow{4}{*}{April 2021}      & \textbf{AutoML4ETC} & \textbf{87.18} & \textbf{87.83} & \textbf{87.18} & \textbf{90.15} & \textbf{121,544} & \textbf{118,984} \\ \cline{2-8} 
                                 & UWOrange-H              & 84.59          & 86.62          & 84.59          & 90.95          & 7,588,360        & 7,588,360        \\ \cline{2-8} 
                                 & UCDavis CNN        & 82.17          & 82.3           & 82.17          & 82.87          & 6,507,016        & 6,507,016        \\ \cline{2-8} 
                                 & DeepPacket CNN      & 82.98          & 82.98          & 82.98          & 83.02          & 9,960,732        & 9,960,732       
                                 \\ \cline{2-8} 
                                 & E2E CNN         & 76.30         & 76.45          & 76.30         & 77.00          & 11,202,440       & 11,202,440       
                                 \\ \hline
\multirow{4}{*}{May 2021}        & \textbf{AutoML4ETC} & \textbf{82.86} & \textbf{84.03} & \textbf{82.85} & \textbf{88.2}  & \textbf{111,560} & \textbf{109,256} \\ \cline{2-8} 
                                 & UWOrange-H              & 79.00             & 80.00             & 79.00             & 83.86          & 7,588,360        & 7,588,360        \\ \cline{2-8} 
                                 & UCDavis CNN        & 79.29          & 79.38          & 79.29          & 79.56          & 6,507,016        & 6,507,016        \\ \cline{2-8} 
                                 & DeepPacket CNN      & 79.54          & 79.59          & 79.54          & 79.66          & 9,960,732        & 9,960,732        
                                 \\ \cline{2-8} 
                                 & E2E CNN         & 76.32         & 76.28          & 76.32          & 76.28         & 11,202,440       & 11,202,440      
                                 \\ \hline
\multirow{4}{*}{June 2021}       & \textbf{AutoML4ETC} & \textbf{87.04} & \textbf{87.69} & \textbf{87.04} & \textbf{89.86} & \textbf{211,144} & \textbf{207,048} \\ \cline{2-8} 
                                 & UWOrange-H              & 86.27          & 86.89          & 86.27          & 89.07          & 7,588,360        & 7,588,360        \\ \cline{2-8} 
                                 & UCDavis CNN        & 83.27          & 83.27          & 83.27          & 83.31          & 6,507,016        & 6,507,016        \\ \cline{2-8} 
                                 & DeepPacket CNN      & 83.37          & 83.37          & 83.37          & 83.41          & 9,960,732        & 9,960,732      
                                 \\ \cline{2-8} 
                                 & E2E CNN         & 75.66          & 75.70          & 75.66          & 75.92          & 11,202,440        & 11,202,440    
                                 \\ \hline
\multirow{4}{*}{QUIC - May 2021} & \textbf{AutoML4ETC} & \textbf{84.22} & \textbf{87.24} & \textbf{84.22} & \textbf{92.21} & \textbf{111,302} & \textbf{108,998} \\ \cline{2-8} 
                                 & UWOrange-H              & 71.48          & 78.23          & 71.48          & 91             & 2,672,902        & 2,672,902        \\ \cline{2-8} 
                                 & UCDavis CNN        & 58.13          & 67.77          & 58.13          & 86.76          & 1,263,878        & 1,263,878        \\ \cline{2-8} 
                                 & DeepPacket CNN      & 28.46          & 38.70          & 28.46          & 87.95          & 1,768,474        & 1,768,474
                                 \\ \cline{2-8} 
                                 & E2E CNN          & 55.42          & 65.92         & 55.42          & 87.92          & 1,894,278      & 1,894,278      
                                 \\ \hline
\multirow{4}{*}{ISCX2016 \cite{icxdataset}} & \textbf{AutoML4ETC} & \textbf{94.35} & \textbf{94.40} & \textbf{94.35} & \textbf{94.87} & \textbf{226,508} & \textbf{222,412} \\ \cline{2-8} 
                                 & UWOrange-H              &     92.81      &     92.56      &     92.81     &       94.87       &   6,360,076     &     6,360,076    \\ \cline{2-8} 
                                 & UCDavis CNN        &      93.78     &      93.82     &      93.78     &     94.01     &    6,360,076     &    6,360,076     \\ \cline{2-8} 
                                 & DeepPacket CNN      &      92.24     &     92.04      &     92.24     &      94.32    &    9,730,848     &    9,730,848   
                                 \\ \cline{2-8} 
                                 & E2E CNN         &    92.48       &      92.47    &    92.48     &       93.30   &     10,944,396   &    10,944,396    
                                 \\ \hline
\end{tabular}%

\end{table*}

\begin{table*}[tb!]
\caption{AutoML4ETC versus the state-of-the-art models in application-level classification}
\label{tab:automl_vs_baselines_rawbytes_applicationlevel}
\centering
\scriptsize
\begin{tabular}{|c|c|c|c|c|c|c|c|}
\hline
Dataset &
  Model &
  \begin{tabular}[c]{@{}c@{}}Accuracy\\ (\%)\end{tabular} &
  \begin{tabular}[c]{@{}c@{}}W Avg.\\  F-1 score (\%)\end{tabular} &
  \begin{tabular}[c]{@{}c@{}}W Avg. \\ recall (\%)\end{tabular} &
  \begin{tabular}[c]{@{}c@{}}W Avg. \\ precision (\%)\end{tabular} &
  Total parameters &
  Trainable parameters \\ \hline\hline
\multirow{4}{*}{July 2019}       & \textbf{AutoML4ETC}     & \textbf{98.48} & \textbf{98.48} & \textbf{98.48} & \textbf{98.5}  & \textbf{271,699} & \textbf{266,835} \\ \cline{2-8} 
                                 & UWOrange-H              & 96.05          & 96.19          & 96.05          & 96.6           & 7,588,360        & 7,588,360        \\ \cline{2-8} 
                                 & UCDavis CNN        & 93.3           & 93.26          & 93.3           & 93.27          & 6,507,016        & 6,507,016        \\ \cline{2-8} 
                                 & DeepPacket CNN      & 91.99           & 91.97           & 91.99           & 91.99           & 9,962,151        & 9,962,151        
                                 \\ \cline{2-8} 
                                 & E2E CNN         & 85.99           & 86.25           & 85.99           & 86.95           & 11,213,715       & 11,213,715     
                                 \\ \hline
\multirow{4}{*}{September 2021}  & \textbf{AutoML4ETC}     & \textbf{88.76} & \textbf{90.16} & \textbf{88.76} & \textbf{94.22} & \textbf{307,027} & \textbf{301,651} \\ \cline{2-8} 
                                 & UWOrange-H              & 84.75          & 86.05          & 84.75          & 90.66          & 7,588,360        & 7,588,360        \\ \cline{2-8} 
                                 & UCDavis CNN        & 88.25          & 88.16          & 88.25          & 88.19          & 6,507,016        & 6,507,016        \\ \cline{2-8} 
                                 & DeepPacket CNN      & 89.83           & 89.94           & 89.83           & 90.27           & 9,962,151        & 9,962,151      
                                 \\ \cline{2-8} 
                                 & E2E CNN         & 85.24           & 85.40           & 85.24           & 85.67           & 11,213,715        & 11,213,715      
                                 \\ \hline
\multirow{4}{*}{April 2021}      & \textbf{AutoML4ETC}     & \textbf{90.76} & \textbf{91.74} & \textbf{90.76} & \textbf{94.5}  & \textbf{236,883} & \textbf{232,531} \\ \cline{2-8} 
                                 & UWOrange-H              & 87.01          & 87.94          & 87.01          & 90.98          & 7,588,360        & 7,588,360        \\ \cline{2-8} 
                                 & UCDavis CNN        & 85.21          & 85.16          & 85.21          & 85.24          & 6,507,016        & 6,507,016        \\ \cline{2-8} 
                                 & DeepPacket CNN      & 85.76           & 85.87           & 85.76           & 86.19           & 9,962,151        & 9,962,151       
                                 \\ \cline{2-8} 
                                 & E2E CNN          & 77.85          & 78.48           & 77.85           & 80.03           & 11,213,715       & 11,213,715  
                                 \\ \hline
\multirow{4}{*}{May 2021}        & \textbf{AutoML4ETC}     & \textbf{84.28} & \textbf{85.64} & \textbf{84.28} & \textbf{90.62} & \textbf{193,363} & \textbf{189,779} \\ \cline{2-8} 
                                 & UWOrange-H              & 74.08          & 80.77          & 74.08          & 90.87          & 7,588,360        & 7,588,360        \\ \cline{2-8} 
                                 & UCDavis CNN        & 80.4           & 80.34          & 80.4           & 80.62          & 6,507,016        & 6,507,016        \\ \cline{2-8} 
                                 & DeepPacket CNN      & 81.17           & 81.29           & 81.17           & 81.62           & 9,962,151        & 9,962,151        
                                 \\ \cline{2-8} 
                                 & E2E CNN          & 82.25          & 82.54           & 82.25           & 83.10           & 11,213,715        & 11,213,715      
                                 \\ \hline
\multirow{4}{*}{June 2021}       & \textbf{AutoML4ETC}     & \textbf{89.53} & \textbf{90.79} & \textbf{89.53} & \textbf{94.27} & \textbf{184,403} & \textbf{181,075} \\ \cline{2-8} 
                                 & UWOrange-H              & 87.24          & 88.45          & 87.24          & 92.06          & 7,588,360        & 7,588,360        \\ \cline{2-8} 
                                 & UCDavis CNN        & 85.17          & 84.97          & 85.17          & 85.02          & 6,507,016        & 6,507,016        \\ \cline{2-8} 
                                 & DeepPacket CNN      & 85.06           & 85.05           & 85.06           & 85.15           & 9,962,151        & 9,962,151       
                                 \\ \cline{2-8} 
                                 & E2E CNN          & 81.15           & 81.18           & 81.15          & 81.32          & 11,213,715       & 11,213,715       
                                 \\ \hline
\multirow{4}{*}{QUIC - May 2021} & \textbf{AutoML4ETC} & \textbf{86.30} & \textbf{85.24} & \textbf{86.30} & \textbf{85.03} & \textbf{191,944} & \textbf{188,360} \\ \cline{2-8} 
                                 & UWOrange-H             & 86.35          & 87.31          & 86.35          & 89.02          & 2,673,160        & 2,673,160        \\ \cline{2-8} 
                                 & UCDavis CNN       & 51.52          & 53.22          & 51.52          & 60.72          & 1,264,136        & 1,264,136        \\ \cline{2-8} 
                                 & DeepPacket CNN      & 26.21           & 14.82           & 26.21           & 15.54           & 1,768,732        & 1,768,732        
                                 \\ \cline{2-8} 
                                 & E2E CNN         & 39.21           & 40.08           & 39.21          & 45.53          & 1,896,328        & 1,896,328       
                                 \\ \hline
\multirow{4}{*}{QUIC - UCDavis \cite{quic-ucdavis}}  & \textbf{AutoML4ETC} & \textbf{100}   & \textbf{100}   & \textbf{100}   & \textbf{100}   & \textbf{130,117} & \textbf{127,301} \\ \cline{2-8} 
                                 & UWOrange-H             & 99.25          & 99.24          & 99.25          & 99.25          & 12,798,085       & 12,798,085       \\ \cline{2-8} 
                                 & UCDavis CNN       & 97.30          & 97.29          & 97.30          & 97.30          & 12,798,085       & 12,798,085       \\ \cline{2-8} 
                                 & DeepPacket CNN      & 98.42           & 98.42           & 98.42           & 98.44           & 19,790,745       & 19,790,745    
                                 \\ \cline{2-8} 
                                 & E2E CNN          & 98.95           & 98.94           & 98.95          & 98.95           & 22,406,021       & 22,406,021   
                                 \\ \hline
\multirow{4}{*}{ISCX2016 \cite{icxdataset}}  & \textbf{AutoML4ETC} & \textbf{92.67}   & \textbf{92.85}   & \textbf{92.67}   & \textbf{93.73}   & \textbf{192,081} & \textbf{188,497} \\ \cline{2-8} 
                                 & UWOrange-H             &     89.15      &     89.14      &     89.15      &      90.93    &    6,360,721    &      6,360,721  \\ \cline{2-8} 
                                 & UCDavis CNN       & 88.19         & 88.36          & 88.19          & 89.05          & 6,360,721       & 6,360,721       \\ \cline{2-8} 
                                 & DeepPacket CNN      & 89.21           & 89.43            & 89.21            & 90.01            & 9,731,493      & 9,731,493     
                                 \\ \cline{2-8} 
                                 & E2E CNN         & 89.30           & 89.14           &89.30           & 90.52           & 10,949,521      & 10,949,521     
                                 \\ \hline
\end{tabular}%
\vspace{-10pt}
\end{table*}

\subsection{AutoML4ETC versus State of the art}\label{ss:automlresults}

In the previous sections, we concluded that using RL as the search algorithm, 40 epochs for child model training, and 100 trials would be our choices for  AutoML4ETC. We now compare the AutoML4ETC-generated model to other state-of-the-art architectures for ETC \blue{(\cf Section \ref{s:baseline})}. \blue{We use the same batch size for all models and also use the input features suggested in \cite{Baseline} (\ie TLS handshake header) for AutoML4ETC.}

Table~\ref{tab:automl_vs_baselines_rawbytes} presents the performance of the AutoML4ETC approach compared with other state-of-the-art ETC models in service-level classification. It is evident that AutoML4ETC  outperforms the state-of-the-art models across all the datasets. In fact, the AutoML4ETC-generated model is $\sim$1 to 60.1\% more accurate than the state of the art. Moreover, it is simpler and lighter, with over 50 times fewer parameters than the state-of-the-art models on average. Table~\ref{tab:automl_vs_baselines_rawbytes_applicationlevel} presents the same results for application-level classification. 

\blue{Interestingly, on the synthetic QUIC - UCDavis dataset \cite{quic-ucdavis}, AutoML4ETC achieves 100\% accuracy recording the highest achievable accuracy with around 100 times fewer parameters than the state-of-the-art models. The detailed architecture is depicted in Appendix Fig. \ref{fig:UCDavisQUIC-AutoML4ETC} as an example. The architecture found by AutoML4ETC is totally different from the hand-crafted state-of-the-art models, which further reinforces the benefit of AutoML4ETC.}

\section{Conclusion \& Future Works} \label{sec:conclude}

In this paper, we introduce~\textit{AutoML4ETC} a novel and fully automated neural architecture search tool for the early classification of encrypted traffic. AutoML4ETC is comprised of a powerful neural architecture search space tailored for raw packet byte-based classification, whose building blocks were carefully selected based on an exhaustive literature review along with our experience in designing and experimenting with a state-of-the-art encrypted traffic classifier. The neural architecture search space is supplemented with efficient search algorithms and training strategies, whose evaluation is reported on several public benchmark and Orange mobile network traffic datasets. We showcase that AutoML4ETC-generated architectures are significantly more light-weight, more effective, and better performing than state-of-the-art ETC models on diverse datasets. 

In recent years, Attention Networks \cite{vaswani2017attention} and Transformer-based \cite{ETC_SIGKDD_Transformer} architectures have shown good performance in ETC. However, their unique architectures with no obvious repetitive structure, pose challenges for AutoML. In the future, we will investigate the possibility of including Attention Networks and Transformers in the NAS search space. 
We will also investigate more efficient ways for searching better performing neural architectures, for example, by using one-shot approaches \cite{Bender2018UnderstandingAS}.

\bibliographystyle{IEEEtran}
\bibliography{bibliography}

\begin{thebibliography}{10}
\providecommand{\url}[1]{#1}
\csname url@samestyle\endcsname
\providecommand{\newblock}{\relax}
\providecommand{\bibinfo}[2]{#2}
\providecommand{\BIBentrySTDinterwordspacing}{\spaceskip=0pt\relax}
\providecommand{\BIBentryALTinterwordstretchfactor}{4}
\providecommand{\BIBentryALTinterwordspacing}{\spaceskip=\fontdimen2\font plus
\BIBentryALTinterwordstretchfactor\fontdimen3\font minus \fontdimen4\font\relax}
\providecommand{\BIBforeignlanguage}[2]{{%
\expandafter\ifx\csname l@#1\endcsname\relax
\typeout{** WARNING: IEEEtran.bst: No hyphenation pattern has been}%
\typeout{** loaded for the language `#1'. Using the pattern for}%
\typeout{** the default language instead.}%
\else
\language=\csname l@#1\endcsname
\fi
#2}}
\providecommand{\BIBdecl}{\relax}
\BIBdecl

\bibitem{aceto2019mobile}
G.~Aceto, D.~Ciuonzo, A.~Montieri, and A.~Pescap{\'e}, ``Mobile encrypted traffic classification using deep learning: Experimental evaluation, lessons learned, and challenges,'' \emph{IEEE Transactions on Network and Service Management}, vol.~16, no.~2, pp. 445--458, 2019.

\bibitem{lotfollahi2020deep}
M.~Lotfollahi, M.~Jafari~Siavoshani, R.~Shirali Hossein~Zade, and M.~Saberian, ``Deep packet: A novel approach for encrypted traffic classification using deep learning,'' \emph{Soft Computing}, vol.~24, no.~3, pp. 1999--2012, 2020.

\bibitem{liu2019fs}
C.~Liu, L.~He, G.~Xiong, Z.~Cao, and Z.~Li, ``Fs-net: A flow sequence network for encrypted traffic classification,'' in \emph{IEEE Conference On Computer Communications (INFOCOM)}, 2019, pp. 1171--1179.

\bibitem{seq2img}
Z.~Chen, K.~He, J.~Li, and Y.~Geng, ``Seq2img: A sequence-to-image based approach towards ip traffic classification using convolutional neural networks,'' in \emph{IEEE International conference on big data (big data)}, 2017, pp. 1271--1276.

\bibitem{CACM}
\BIBentryALTinterwordspacing
I.~Akbari, M.~A. Salahuddin, L.~Ven, N.~Limam, R.~Boutaba, B.~Mathieu, S.~Moteau, and S.~Tuffin, ``Traffic classification in an increasingly encrypted web,'' \emph{Commun. ACM}, vol.~65, no.~10, p. 75–83, sep 2022. [Online]. Available: \url{https://doi.org/10.1145/3559439}
\BIBentrySTDinterwordspacing

\bibitem{deeptor}
A.~Khajehpour, F.~Zandi \emph{et~al.}, ``Deep inside tor: Exploring website fingerprinting attacks on tor traffic in realistic settings,'' in \emph{International Conf. on Computer and Knowledge Engineering}, 2022, pp. 148--156.

\bibitem{FSTC}
N.~Malekghaini, H.~Tsang, M.~A. Salahuddin, N.~Limam, and R.~Boutaba, ``Fstc: Dynamic category adaptation for encrypted network traffic classification,'' in \emph{2023 IFIP Networking Conference (IFIP Networking)}, 2023, pp. 1--9.

\bibitem{Baseline}
I.~Akbari, M.~A. Salahuddin, L.~Ven, N.~Limam, R.~Boutaba, B.~Mathieu, S.~Moteau, and S.~Tuffin, ``A look behind the curtain: traffic classification in an increasingly encrypted web,'' \emph{Proceedings of the ACM on Measurement and Analysis of Computing Systems}, vol.~5, no.~1, pp. 1--26, 2021.

\bibitem{IFIPPaper}
N.~Malekghaini, E.~Akbari, M.~A. Salahuddin, N.~Limam, R.~Boutaba, B.~Mathieu, S.~Moteau, and S.~Tuffin, ``Data drift in dl: Lessons learned from encrypted traffic classification,'' in \emph{IFIP Networking Conference (IFIP Networking)}, 2022, pp. 1--9.

\bibitem{COMNETPaper}
N.~Malekghaini, E.~Akbari \emph{et~al.}, ``Deep learning for encrypted traffic classification in the face of data drift: An empirical study,'' \emph{Computer Networks}, vol. 225, p. 109648, 2023.

\bibitem{scholarpedia}
T.~Schaul and J.~Schmidhuber, ``Metalearning,'' \emph{Scholarpedia}, vol.~5, no.~6, p. 4650, 2010.

\bibitem{metalearnclassic1}
S.~Hochreiter, A.~S. Younger, and P.~R. Conwell, ``Learning to learn using gradient descent,'' in \emph{International conference on artificial neural networks}.\hskip 1em plus 0.5em minus 0.4em\relax Springer, 2001, pp. 87--94.

\bibitem{NASsurvey}
T.~Elsken, J.~H. Metzen, and F.~Hutter, ``Neural architecture search: A survey,'' \emph{The Journal of Machine Learning Research}, vol.~20, no.~1, pp. 1997--2017, 2019.

\bibitem{NAS}
\BIBentryALTinterwordspacing
B.~Zoph and Q.~V. Le, ``Neural architecture search with reinforcement learning,'' in \emph{5th International Conference on Learning Representations, {ICLR} 2017, Toulon, France, April 24-26, 2017, Conference Track Proceedings}.\hskip 1em plus 0.5em minus 0.4em\relax OpenReview.net, 2017. [Online]. Available: \url{https://openreview.net/forum?id=r1Ue8Hcxg}
\BIBentrySTDinterwordspacing

\bibitem{automl_classic}
M.~Feurer, A.~Klein, K.~Eggensperger, J.~Springenberg, M.~Blum, and F.~Hutter, ``Efficient and robust automated machine learning,'' in \emph{Advances in Neural Information Processing Systems}, C.~Cortes, N.~Lawrence, D.~Lee, M.~Sugiyama, and R.~Garnett, Eds., vol.~28.\hskip 1em plus 0.5em minus 0.4em\relax Curran Associates, Inc., 2015.

\bibitem{Resnet}
K.~He, X.~Zhang, S.~Ren, and J.~Sun, ``Deep residual learning for image recognition,'' in \emph{IEEE conference on computer vision and pattern recognition}, 2016, pp. 770--778.

\bibitem{cifar10}
A.~Krizhevsky, G.~Hinton \emph{et~al.}, ``Learning multiple layers of features from tiny images,'' 2009.

\bibitem{NAStoENAS}
B.~Zoph, V.~Vasudevan, J.~Shlens, and Q.~V. Le, ``Learning transferable architectures for scalable image recognition,'' in \emph{IEEE conference on computer vision and pattern recognition}, 2018, pp. 8697--8710.

\bibitem{deng2009imagenet}
J.~Deng, W.~Dong, R.~Socher, L.-J. Li, K.~Li, and L.~Fei-Fei, ``Imagenet: A large-scale hierarchical image database,'' in \emph{2009 IEEE Conference on Computer Vision and Pattern Recognition}, 2009, pp. 248--255.

\bibitem{ENAS}
H.~Pham, M.~Guan, B.~Zoph, Q.~Le, and J.~Dean, ``Efficient neural architecture search via parameters sharing,'' in \emph{International conference on machine learning}.\hskip 1em plus 0.5em minus 0.4em\relax PMLR, 2018, pp. 4095--4104.

\bibitem{multitask}
\BIBentryALTinterwordspacing
S.~Ruder, ``An overview of multi-task learning in deep neural networks,'' 2017. [Online]. Available: \url{https://arxiv.org/abs/1706.05098}
\BIBentrySTDinterwordspacing

\bibitem{MCTSTheory}
L.~Kocsis and C.~Szepesv{\'a}ri, ``Bandit based monte-carlo planning,'' in \emph{European conference on machine learning}.\hskip 1em plus 0.5em minus 0.4em\relax Springer, 2006, pp. 282--293.

\bibitem{MCTS}
L.~Wang, Y.~Zhao, Y.~Jinnai, Y.~Tian, and R.~Fonseca, ``Neural architecture search using deep neural networks and monte carlo tree search,'' in \emph{AAAI Conference on Artificial Intelligence}, vol.~34, no.~06, 2020, pp. 9983--9991.

\bibitem{EAlarge}
E.~Real, S.~Moore, A.~Selle, S.~Saxena, Y.~L. Suematsu, J.~Tan, Q.~V. Le, and A.~Kurakin, ``Large-scale evolution of image classifiers,'' in \emph{International Conference on Machine Learning}.\hskip 1em plus 0.5em minus 0.4em\relax PMLR, 2017, pp. 2902--2911.

\bibitem{EA}
E.~Real, A.~Aggarwal, Y.~Huang, and Q.~V. Le, ``Regularized evolution for image classifier architecture search,'' in \emph{{AAAI} conference on artificial intelligence}, vol.~33, no.~01, 2019, pp. 4780--4789.

\bibitem{herrmann2009website}
D.~Herrmann, R.~Wendolsky, and H.~Federrath, ``Website fingerprinting: attacking popular privacy enhancing technologies with the multinomial na{\"\i}ve-bayes classifier,'' in \emph{Proceedings of the 2009 ACM workshop on Cloud computing security}, 2009, pp. 31--42.

\bibitem{armitage_iman}
N.~Williams, S.~Zander, and G.~Armitage, ``A preliminary performance comparison of five machine learning algorithms for practical ip traffic flow classification,'' \emph{ACM SIGCOMM Computer Communication Review}, vol.~36, no.~5, pp. 5--16, 2006.

\bibitem{classicalMLsurveyraouf}
R.~Boutaba, M.~A. Salahuddin, N.~Limam, S.~Ayoubi, N.~Shahriar, F.~Estrada-Solano, and O.~M. Caicedo, ``A comprehensive survey on machine learning for networking: evolution, applications and research opportunities,'' \emph{Journal of Internet Services and Applications}, vol.~9, no.~1, pp. 1--99, 2018.

\bibitem{wang2017end}
W.~Wang, M.~Zhu, J.~Wang, X.~Zeng, and Z.~Yang, ``End-to-end encrypted traffic classification with one-dimensional convolution neural networks,'' in \emph{2017 IEEE international conference on intelligence and security informatics (ISI)}.\hskip 1em plus 0.5em minus 0.4em\relax IEEE, 2017, pp. 43--48.

\bibitem{FlowPic}
T.~Shapira and Y.~Shavitt, ``Flowpic: Encrypted internet traffic classification is as easy as image recognition,'' in \emph{IEEE INFOCOM 2019 - IEEE Conference on Computer Communications Workshops (INFOCOM WKSHPS)}, 2019, pp. 680--687.

\bibitem{fsnet}
C.~Liu, L.~He, G.~Xiong, Z.~Cao, and Z.~Li, ``Fs-net: A flow sequence network for encrypted traffic classification,'' in \emph{IEEE INFOCOM 2019-IEEE Conference On Computer Communications}.\hskip 1em plus 0.5em minus 0.4em\relax IEEE, 2019, pp. 1171--1179.

\bibitem{aceto2019mimetic}
G.~Aceto, D.~Ciuonzo, A.~Montieri, and A.~Pescap{\`e}, ``Mimetic: Mobile encrypted traffic classification using multimodal deep learning,'' \emph{Computer networks}, vol. 165, p. 106944, 2019.

\bibitem{et-bert}
\BIBentryALTinterwordspacing
X.~Lin, G.~Xiong, G.~Gou, Z.~Li, J.~Shi, and J.~Yu, ``Et-bert: A contextualized datagram representation with pre-training transformers for encrypted traffic classification,'' in \emph{Proceedings of the ACM Web Conference 2022}, ser. WWW '22.\hskip 1em plus 0.5em minus 0.4em\relax New York, NY, USA: Association for Computing Machinery, 2022, p. 633–642. [Online]. Available: \url{https://doi.org/10.1145/3485447.3512217}
\BIBentrySTDinterwordspacing

\bibitem{largescaleUCDavis}
S.~Rezaei, B.~Kroencke, and X.~Liu, ``Large-scale mobile app identification using deep learning,'' \emph{IEEE Access}, vol.~8, pp. 348--362, 2019.

\bibitem{iscx-vpn}
G.~Draper-Gil, A.~H. Lashkari, M.~S.~I. Mamun, and A.~A. Ghorbani, ``Characterization of encrypted and vpn traffic using time-related,'' in \emph{Proceedings of the 2nd international conference on information systems security and privacy (ICISSP)}, 2016, pp. 407--414.

\bibitem{iscx-tor}
A.~H. Lashkari, G.~D. Gil, M.~S.~I. Mamun, and A.~A. Ghorbani, ``Characterization of tor traffic using time based features,'' in \emph{International Conference on Information Systems Security and Privacy}, vol.~2.\hskip 1em plus 0.5em minus 0.4em\relax SciTePress, 2017, pp. 253--262.

\bibitem{quic-ucdavis}
S.~Rezaei and X.~Liu, ``How to achieve high classification accuracy with just a few labels: A semi-supervised approach using sampled packets,'' \emph{arXiv preprint arXiv:1812.09761}, 2018.

\bibitem{rossilessonslearned}
L.~Yang, A.~Finamore, F.~Jun, and D.~Rossi, ``Deep learning and zero-day traffic classification: Lessons learned from a commercial-grade dataset,'' \emph{IEEE Transactions on Network and Service Management}, vol.~18, no.~4, pp. 4103--4118, 2021.

\bibitem{nprint}
J.~Holland, P.~Schmitt, N.~Feamster, and P.~Mittal, ``New directions in automated traffic analysis,'' in \emph{Proceedings of the 2021 ACM SIGSAC Conference on Computer and Communications Security}, 2021, pp. 3366--3383.

\bibitem{ggfast}
J.~Piet, D.~Nwoji, and V.~Paxson, ``Ggfast: Automating generation of flexible network traffic classifiers,'' in \emph{Proceedings of the ACM SIGCOMM 2023 Conference}, 2023, pp. 850--866.

\bibitem{autogluon}
N.~Erickson, J.~Mueller, A.~Shirkov, H.~Zhang, P.~Larroy, M.~Li, and A.~Smola, ``Autogluon-tabular: Robust and accurate automl for structured data,'' \emph{arXiv preprint arXiv:2003.06505}, 2020.

\bibitem{mljar}
\BIBentryALTinterwordspacing
A.~P\l{}o\'{n}ska and P.~P\l{}o\'{n}ski, ``Mljar: State-of-the-art automated machine learning framework for tabular data. version 0.10.3,'' \L{}apy, Poland, 2021. [Online]. Available: \url{https://github.com/mljar/mljar-supervised}
\BIBentrySTDinterwordspacing

\bibitem{automl_classical}
D.~F. Isingizwe, M.~Wang, W.~Liu, D.~Wang, T.~Wu, and J.~Li, ``Analyzing learning-based encrypted malware traffic classification with automl,'' in \emph{IEEE International Conference on Communication Technology (ICCT)}, 2021, pp. 313--322.

\bibitem{E2E}
W.~Wang, M.~Zhu, J.~Wang, X.~Zeng, and Z.~Yang, ``End-to-end encrypted traffic classification with one-dimensional convolution neural networks,'' in \emph{2017 IEEE International Conference on Intelligence and Security Informatics (ISI)}, 2017, pp. 43--48.

\bibitem{tensorflow2015-whitepaper}
\BIBentryALTinterwordspacing
``{TensorFlow}: Large-scale machine learning on heterogeneous systems,'' 2015, software available from tensorflow.org. [Online]. Available: \url{https://www.tensorflow.org/}
\BIBentrySTDinterwordspacing

\bibitem{chollet2015keras}
F.~Chollet \emph{et~al.}, ``Keras,'' \url{https://keras.io}, 2015.

\bibitem{spark}
\BIBentryALTinterwordspacing
M.~Zaharia, R.~S. Xin, P.~Wendell, T.~Das, M.~Armbrust, A.~Dave, X.~Meng, J.~Rosen, S.~Venkataraman, M.~J. Franklin, A.~Ghodsi, J.~Gonzalez, S.~Shenker, and I.~Stoica, ``Apache spark: A unified engine for big data processing,'' \emph{Commun. ACM}, vol.~59, no.~11, p. 56–65, oct 2016. [Online]. Available: \url{https://doi.org/10.1145/2934664}
\BIBentrySTDinterwordspacing

\bibitem{Hypernets}
D.~IO, ``Hypernets,'' \url{https://github.com/DataCanvasIO/Hypernets}, 2021.

\bibitem{ietf-tls-esni-14}
\BIBentryALTinterwordspacing
E.~Rescorla, K.~Oku, N.~Sullivan, and C.~A. Wood, ``{TLS Encrypted Client Hello},'' Internet Engineering Task Force, Internet-Draft draft-ietf-tls-esni-14, Feb. 2022, work in Progress. [Online]. Available: \url{https://datatracker.ietf.org/doc/html/draft-ietf-tls-esni-14}
\BIBentrySTDinterwordspacing

\bibitem{icxdataset}
G.~Draper-Gil, A.~H. Lashkari \emph{et~al.}, ``Characterization of encrypted and vpn traffic using time-related features,'' in \emph{International Conf. on Information Systems Security and Privacy}, 2016, pp. 407--414.

\bibitem{UCT}
P.~Auer, N.~Cesa-Bianchi, and P.~Fischer, ``Finite-time analysis of the multiarmed bandit problem,'' \emph{Machine learning}, vol.~47, no.~2, pp. 235--256, 2002.

\bibitem{Ke2017LightGBMAH}
G.~Ke, Q.~Meng, T.~Finley, T.~Wang, W.~Chen, W.~Ma, Q.~Ye, and T.-Y. Liu, ``Lightgbm: A highly efficient gradient boosting decision tree,'' in \emph{NIPS}, 2017.

\bibitem{vaswani2017attention}
A.~Vaswani, N.~Shazeer, N.~Parmar, J.~Uszkoreit, L.~Jones, A.~N. Gomez, {\L}.~Kaiser, and I.~Polosukhin, ``Attention is all you need,'' in \emph{Advances in neural information processing systems}, 2017, pp. 5998--6008.

\bibitem{ETC_SIGKDD_Transformer}
\BIBentryALTinterwordspacing
R.~Zhao, X.~Deng, Z.~Yan, J.~Ma, Z.~Xue, and Y.~Wang, ``Mt-flowformer: A semi-supervised flow transformer for encrypted traffic classification,'' in \emph{Proceedings of the 28th ACM SIGKDD Conference on Knowledge Discovery and Data Mining}, ser. KDD '22.\hskip 1em plus 0.5em minus 0.4em\relax New York, NY, USA: Association for Computing Machinery, 2022, p. 2576–2584. [Online]. Available: \url{https://doi.org/10.1145/3534678.3539314}
\BIBentrySTDinterwordspacing

\bibitem{Bender2018UnderstandingAS}
G.~Bender, P.-J. Kindermans, B.~Zoph, V.~Vasudevan, and Q.~V. Le, ``Understanding and simplifying one-shot architecture search,'' in \emph{ICML}, 2018.

\end{thebibliography}

\begin{IEEEbiography}[{\includegraphics[width=1.1in,height=1.25in,clip,keepaspectratio]{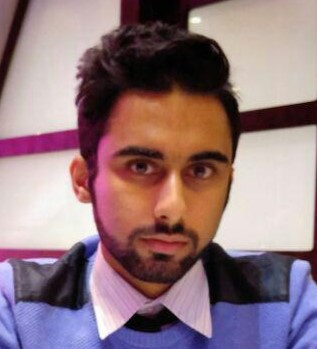}}]%
{Navid Malekghaini}
received his B.Sc. from Sharif University in Computer Engineering and Master's degree in Computer Science at the University of Waterloo. His research revolves around the intersection of machine learning, network security, network management, and cloud computing with a focus on automation and scalability. He is also a reviewer for peer-reviewed journals and conferences. After his Master's, he continued as a Senior ML researcher \& consultant at University of Waterloo, Canada in collaboration with Orange Labs R\&D, France.
\end{IEEEbiography}

\begin{IEEEbiography}[{\includegraphics[width=1.1in,height=1.25in,clip,keepaspectratio]{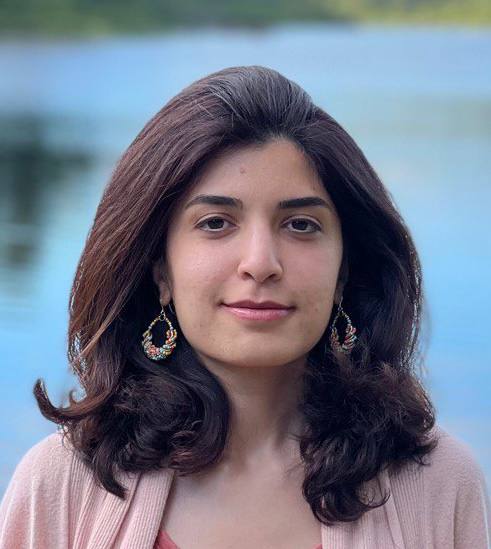}}]%
{Elham Akbari}
received her B.Sc. and M.Sc. degree from Sharif University of Technology. She is currently a Ph.D. student at the University of Waterloo. Her work concerns network traffic classification and the application of few-shot and meta-learning algorithms in encrypted network data analysis.
\end{IEEEbiography}

\begin{IEEEbiography}[{\includegraphics[width=1.1in,height=1.25in,clip,keepaspectratio]{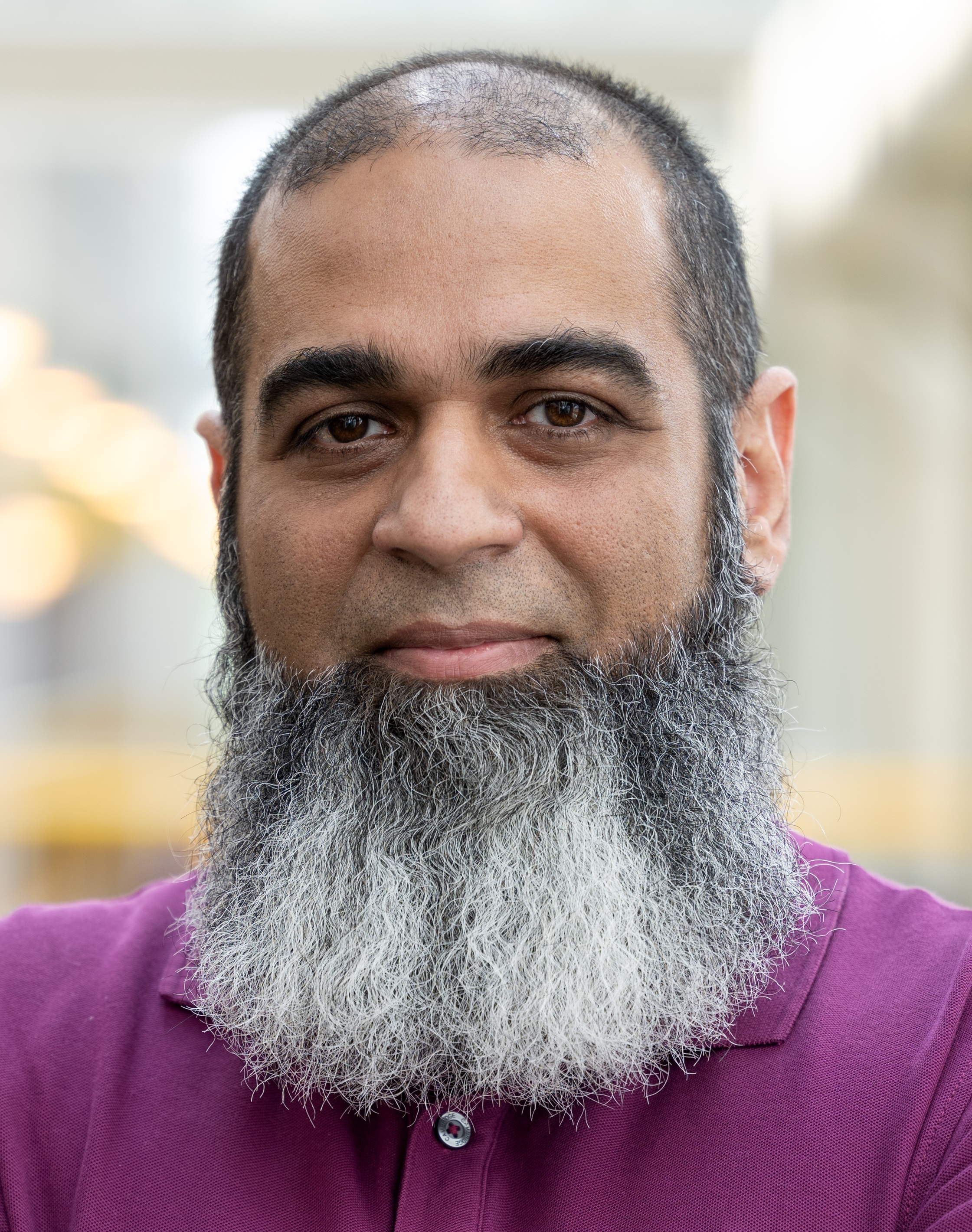}}]%
{Mohammad A. Salahuddin}
received his Ph.D. degree in Computer Science from Western Michigan University in 2014. He is currently a Research Assistant Professor of Computer Science with the University of Waterloo. His research interests include the Internet of Things, content delivery networks, network softwarization, network security, and cognitive network management. His co-authored research publications have received numerous awards, including ACM SIGMETRICS best student paper, IEEE/IFIP NOMS best papers, and IEEE CNOM best paper. He serves on the TPC for international conferences and is a reviewer for various peer-reviewed journals, magazines and conferences.
\end{IEEEbiography}

\begin{IEEEbiography}[{\includegraphics[width=1.1in,height=1.25in,clip,keepaspectratio]{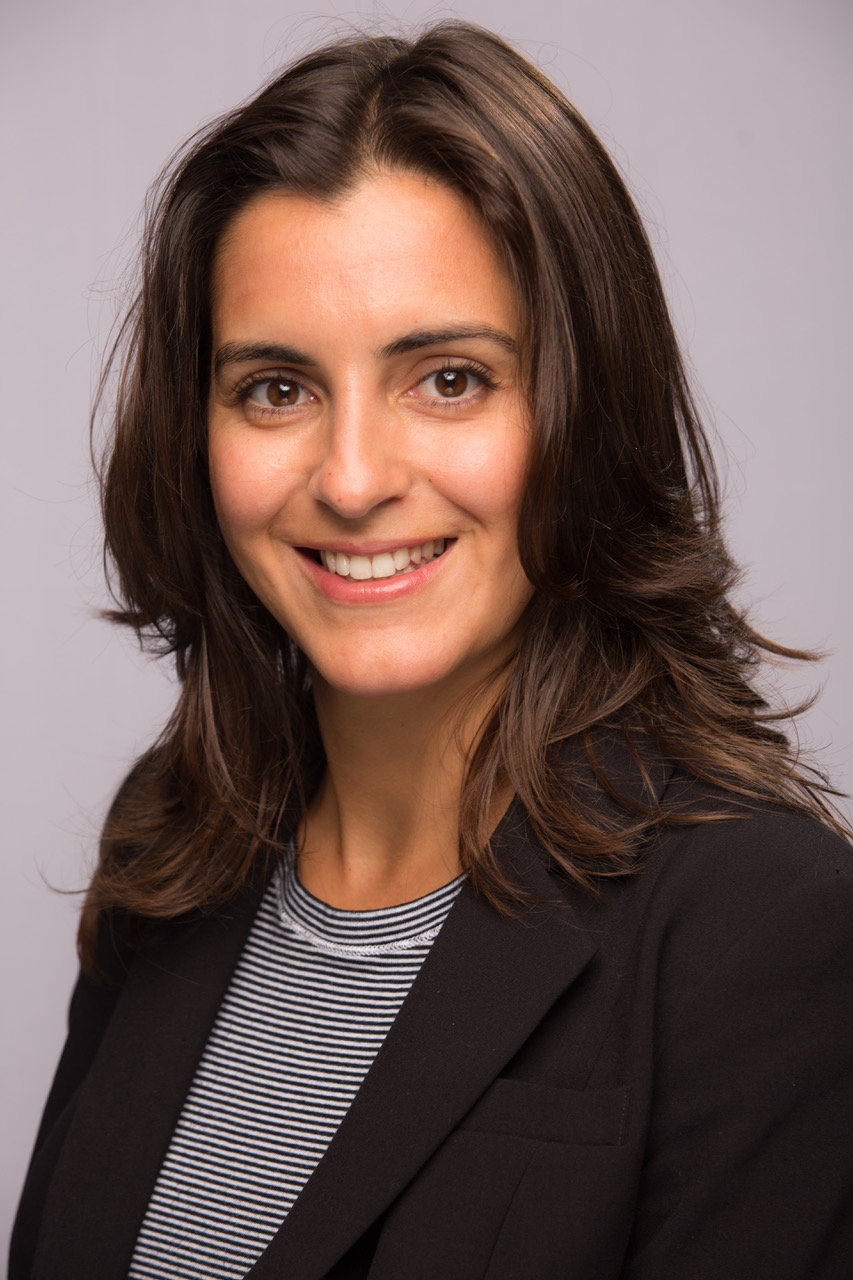}}]%
{Noura Limam}
is a research assistant professor of computer science at the University of Waterloo, Canada. She received the MSc and PhD degrees in computer science from the University Pierre \& Marie Curie (now Sorbonne University), France, in 2002 and 2007, respectively. She is an active researcher and contributor in the area of network and service management. Her current interests revolve around network automation and cognitive network management. She is the chair of the IEEE ComSoc Network Operations and Management Technical Committee, an Associate Editor of the IEEE Communications Magazine, and a Guest Editor of the IEEE Communications Magazine Network Softwarization and Management Series. 
\end{IEEEbiography}

 \begin{IEEEbiography}[{\includegraphics[width=1.1in,height=1.25in,clip,keepaspectratio]{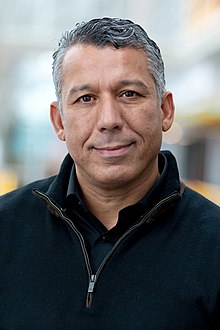}}]%
{Raouf Boutaba}
received the M.Sc. and Ph.D. degrees in computer science from Sorbonne University in 1990 and 1994, respectively. He is currently a University Chair Professor and the Director of the David R. Cheriton School of Computer Science at the University of Waterloo (Canada). He is the founding Editor-in-Chief of the IEEE Transactions on Network and Service Management (2007-2010) and served as the Editor-in-Chief of the IEEE Journal on Selected Areas in Communications (2018-2021). He is a fellow of the IEEE, the Engineering Institute of Canada, the Canadian Academy of Engineering, and the Royal Society of Canada.
\end{IEEEbiography}

 \begin{IEEEbiography}[{\includegraphics[width=1.1in,height=1.25in,clip,keepaspectratio]{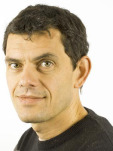}}]%
{Bertrand Mathieu}
received the M.Sc. degree from the University of Marseille, the Ph.D. degree from the Sorbonne University, Paris, and the Habilitation à Diriger des Recherches degree from the University of Rennes. He is a Senior Researcher with Orange Innovation. He contributed to 14 national/European projects and published over 80 papers in international conferences, journals, and books. He is working on programmable networks, QoS and QoE, and new network solutions. He is a member of several conferences’ TPC.
\end{IEEEbiography}

 \begin{IEEEbiography}[{\includegraphics[width=1.1in,height=1.25in,clip,keepaspectratio]{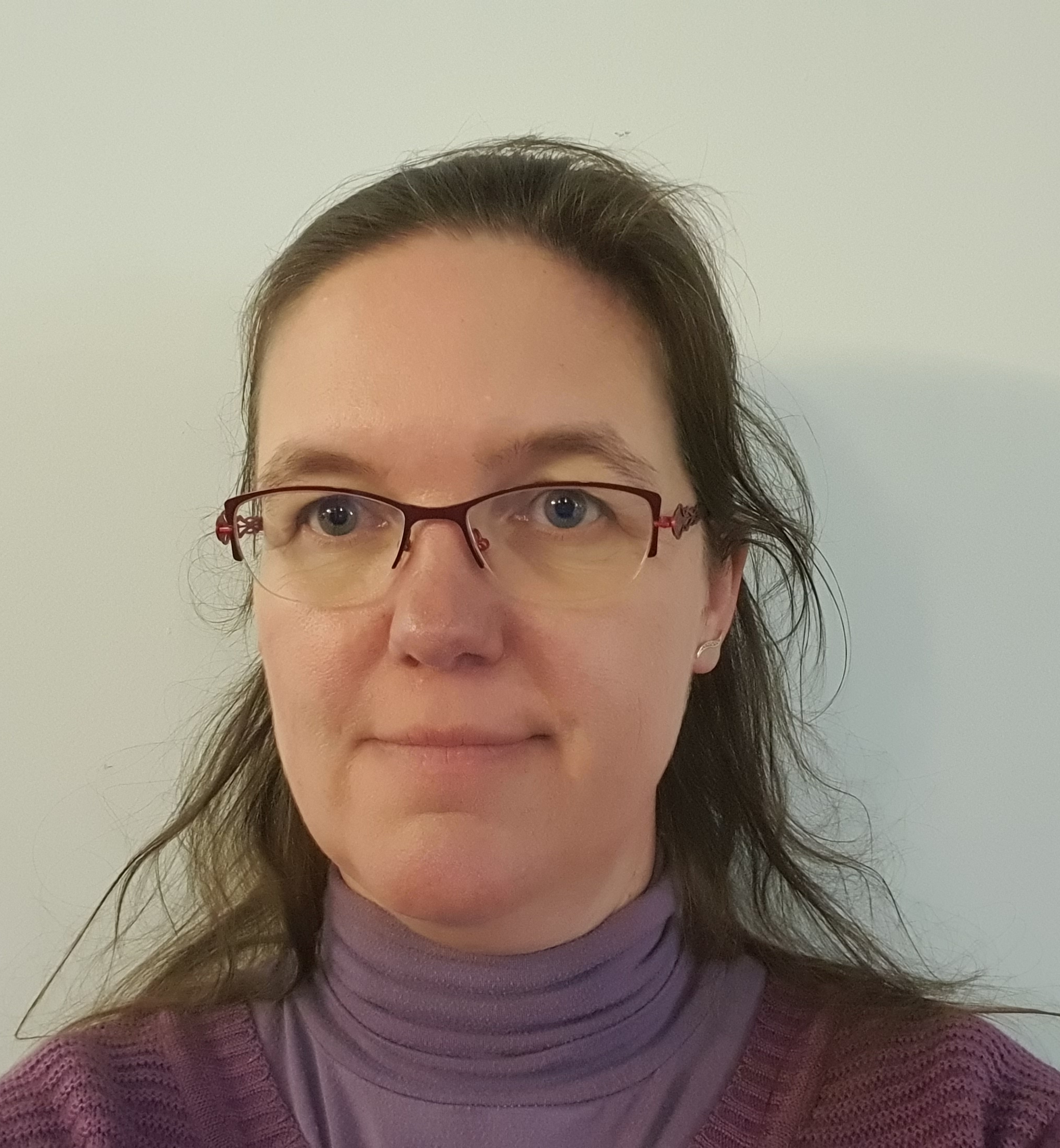}}]%
{Stéphanie Moteau}
is an R\&D engineer at Orange Labs for 18 years recognized as an Orange Expert on Future Networks. She works on the IP metrology domain for various purposes: security, quality of service and quality of experience, Content Delivery Network, traffic classification, etc.   

She participated in many collaborative projects (French and European ones) as a contributor and project leader for Orange. Two years ago, she became a Data Scientist. This gives her new methods to perform her research activity.  
\end{IEEEbiography}

 \begin{IEEEbiography}[{\includegraphics[width=1.1in,height=1.25in,clip,keepaspectratio]{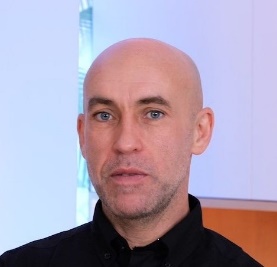}}]%
{Stéphane Tuffin}
is an R\&D engineer at Orange Labs for 20 years recognized as an Orange Expert on Future Networks. He has been deeply involved in the transition from circuit-switched voice to IP and spent several years preparing and supporting IMS deployments in Orange countries.

In 2013, he took the lead of a research project in the field of Web real-time communications which gave birth to a “no backend” offer.

Since 2017, he has been leading a research program on end-to-end network quality which covers network planning, traffic classification, monitoring-troubleshooting the performance of Internet applications, and adapting networks to the need of interactive services. His own research interests are the monitoring of encrypted traffic and low-latency networking. These last years, he has been refocusing research efforts on integrating environmental limits in network management.
\end{IEEEbiography}

\begin{figure*}[htb!]
\centering
\includegraphics[width=1.5\columnwidth]{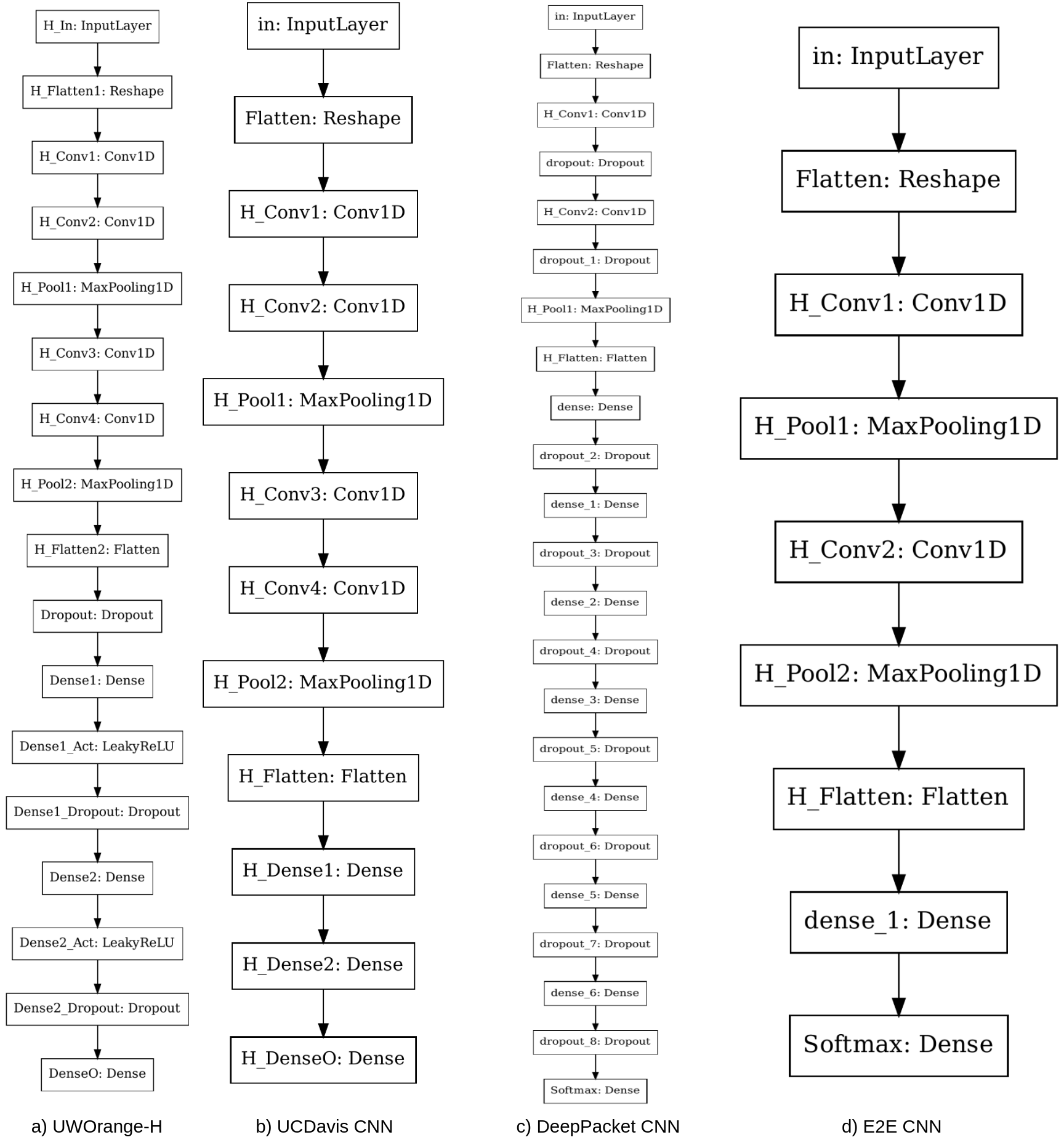}
\caption{\blue{The state-of-the-art ETC baseline model architectures used in this paper}}
\label{fig:sota}

\end{figure*}

\begin{figure*}[th]
\centerline{
\includegraphics[height=1.3\textwidth]{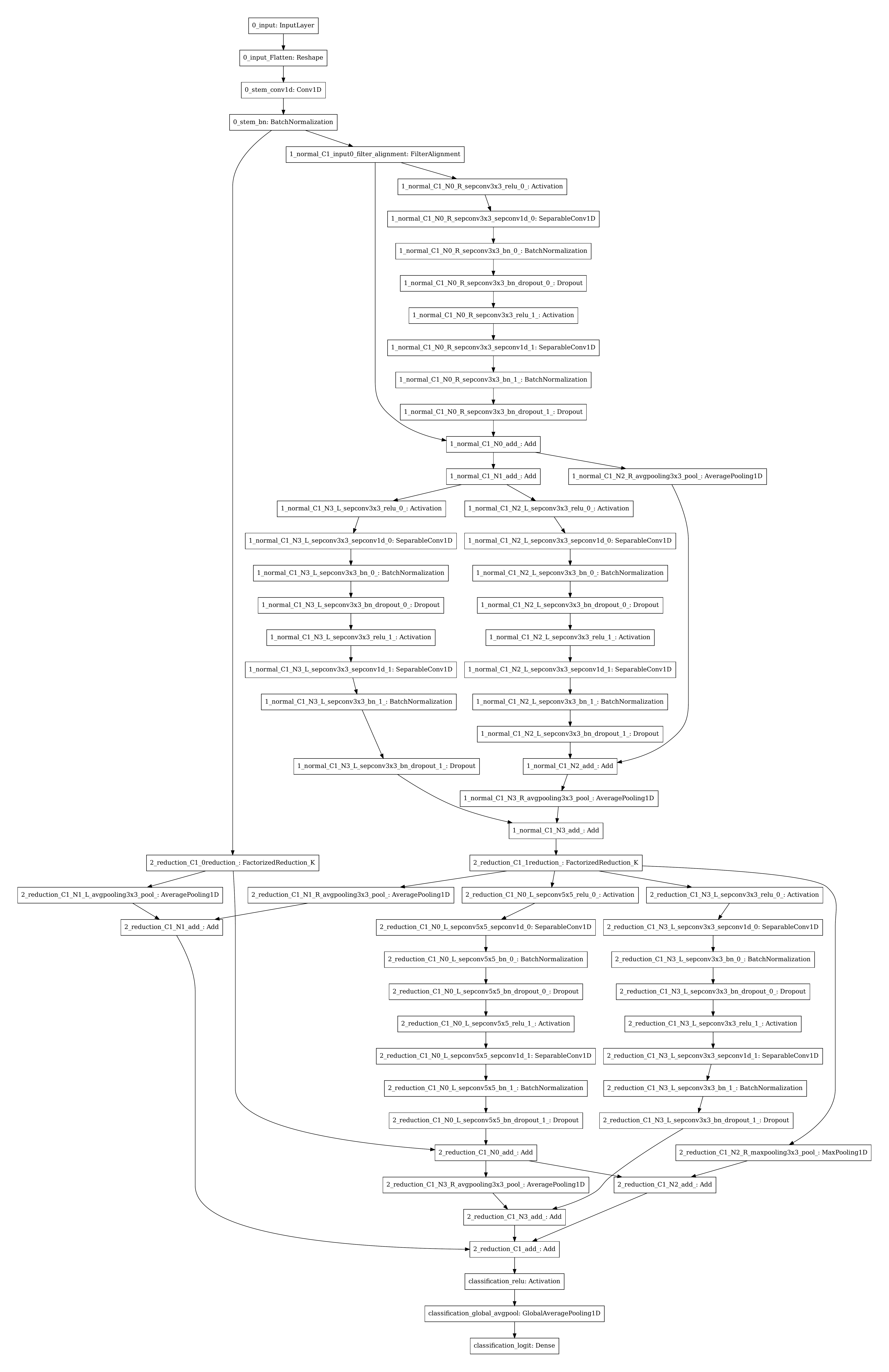}}
\caption{\blue{AutoML4ETC-generated DNN architecture that achieves 100\% accuracy on the QUIC - UCDavis dataset \cite{quic-ucdavis} }}
\label{fig:UCDavisQUIC-AutoML4ETC}

\end{figure*}

\end{document}